\documentclass[12pt]{article}

\usepackage{epsf}
\usepackage[dvips]{graphicx}

\title{Forward-backward correlations between multiplicities in windows separated in azimuth and rapidity}
\author{Vladimir Vechernin\\Saint-Petersburg State University}
\date{}
\begin{document}
\maketitle

\begin{abstract}
The forward-backward (FB) charged particle multiplicity correlations between windows separated in rapidity and
azimuth are analyzed using a model that treats strings as independent identical emitters.
Both the short-range (SR) contribution, originating from the correlation between multiplicities produced
from a single source, and the long-range (LR) contribution, originating from the fluctuation in the number
of sources, are taken into account.
The dependencies of the FB correlation coefficient, $b$, on the windows' rapidity and azimuthal acceptance
and the gaps between these windows are studied and compared with
the preliminary data of ALICE.
The analysis of these dependencies effectively separates the contributions of two above mechanisms.
It is also demonstrated that traditional definitions of FB correlation coefficient $b$ have a strong nonlinear
dependence on the acceptance of windows.  Suitable alternative observables for the future FB correlation
studies are proposed.
The connection between $b$ and the two-particle correlation function, $C_2$, is traced,
as well as its connection to the untriggered di-hadron correlation analysis.
Using a model independent analysis, it is shown that measurement of the FB multiplicity correlations between
two small windows separated in rapidity and azimuth fully determine the two-particle correlation function $C_2$,
even if the particle distribution in rapidity is not uniform.

\ \\
\textbf{Keywords:}
hadronic interactions,
high energy,
soft multiparticle production,
multiplicity correlations

\end{abstract}

\section{Introduction}
\label{Introduction}

%%%%%%%%%%%%%%%%%%%%%
\def\bc{\begin{center}}
\def\ec{\end{center}}
\def\beq{\begin{equation}}
\def\eeq{\end{equation}}
\def\noi{\noindent}
\def\hs#1{\hspace*{#1cm}}
\def\av#1{\langle #1 \rangle}
\def\avL#1{\left\langle #1 \right\rangle}
\def\avo#1{{\av{#1}}}
\def\avr#1#2{\langle {#1} \rangle^{}_{#2}}
\def\pir{{\frac{1}{2\pi}}}
\def\pirr{{(2\pi)^{-2}_{}}}

\def\dpF{{\delta p_{{\rm T}F}^{}}}
\def\dpB{{\delta p_{{\rm T}B}^{}}}
\def\bpF{{\textbf{p}_{F}^{}}}
\def\bpB{{\textbf{p}_{B}^{}}}

\def\nF{{n_F^{}}}
\def\nB{{n_B^{}}}
\def\nFr{{\frac{\nF}{\av{\nF}}}}
\def\nBr{{\frac{\nB}{\av{\nB}}}}
\def\pF{{p_{tF}^{}}}
\def\pB{{p_{tB}^{}}}
\def\mF{{\mu^{}_F}}
\def\mB{{\mu^{}_B}}
\def\mFr{{\frac{\mF}{\av{\mF}}}}
\def\mBr{{\frac{\mB}{\av{\mB}}}}

\def\omu{{\av{\mu}}}
\def\omF{{\av{\mu^{}_F}}}
\def\omB{{\av{\mu^{}_B}}}
\def\omBF{{\av{\mu^{}_F\mu^{}_B}}}
\def\omFB{{\av{\mu^{}_F\mu^{}_B}}}

\def\mo{{\mu^{}_{0}}}
\def\moo{{\mu^{2}_{0}}}
\def\rhoo{{\rho^{}_{0}}}
\def\rhooo{{\rho^{2}_{0}}}
\def\moF{{\mu^{}_{0F}}}
\def\moB{{\mu^{}_{0B}}}

\def\ommF{{\av{\mu_F^2}}}
\def\ommB{{\av{\mu_B^2}}}
\def\omFF{{\av{\mu_F^{}}_{}^2}}
\def\omBB{{\av{\mu_B^{}}_{}^2}}
\def\df{\delta_{F,\sum F_i}}
\def\db{\delta_{B,\sum B_i}}
\def\pp{\prod_{i=1}^N p(B_i,F_i)}
\def\sumn{\sum^n_{i=1}}

\def\oN{\overline{N}}
\def\obr{\overline{b}^{rel}_{}}
\def\bcor{b_{corr}}
\def\bcorr{b_{corr}^{}}
\def\brel{b_{rel}^{}}
\def\brelLR{b_{rel}^{LR}}
\def\bm{\beta_{mod}^{}}
\def\bmp{\beta_{mod}^{'}}
\def\babs{b_{abs}^{}}
\def\br{b_{rel}^{}}
\def\ba{b_{abs}^{}}
\def\bsym{b_{sym}^{}}
\def\brob{\beta_{rob}^{}}
\def\oba{\overline{b}^{abs}_{}}
\def\ar{a^{rel}_{}}
\def\aa{a^{abs}_{}}
\def\dnF{\nF-\av{\nF}}
\def\pc{\!+\!}
\def\mc{\!-\!}
\def\ppc{\!+\!...\!+\!}
\def\yFB{{\eta^{}_{FB}}}
\def\yBF{{\eta^{}_{FB}}}
\def\yF{{\eta^{}_F}}
\def\yB{{\eta^{}_B}}
\def\fFB{{\phi^{}_{FB}}}
\def\fBF{{\phi^{}_{FB}}}
\def\fF{{\phi^{}_F}}
\def\fB{{\phi^{}_B}}

\def\dyf{{d\eta d\phi}}
\def\dyfo{{d\eta_1 d\phi_1}}
\def\dyft{{d\eta_2 d\phi_2}}
\def\dyp{{d\eta'}}
\def\dfp{{d\phi'}}
\def\dyfp{{d\eta' d\phi'}}

\def\Dyy{{(\delta \eta)^2}}
\def\Dy{{\delta \eta}}
\def\Df{{\delta \phi}}

\def\DyF{{\delta \eta^{}_F}}
\def\DyFF{{(\delta \eta^{}_F)^2}}
\def\DyB{{\delta \eta^{}_B}}

\def\DfFF{{(\delta \phi^{}_F)^2}}
\def\DfF{{\delta \phi^{}_F}}
\def\DfB{{\delta \phi^{}_B}}

\def\dfF{{\delta \phi^{}_F}}
\def\dfB{{\delta \phi^{}_B}}

\def\dyF{{\delta \eta^{}_F}}
\def\dyB{{\delta \eta^{}_B}}

\def\DyfFF{{(\delta \eta^{}_F\delta \phi^{}_F)^2}}
\def\DyfF{{\delta \eta^{}_F\delta \phi^{}_F}}
\def\DyfB{{\delta \eta^{}_B\delta \phi^{}_B}}

\def\aFF{{\delta^{2}_F}}
\def\aFFr{{\delta^{-2}_F}}
\def\aF{{\delta^{}_F}}
\def\aB{{\delta^{}_B}}
\def\aFr{{\delta^{-1}_F}}
\def\aBr{{\delta^{-1}_B}}
\def\aW{{\delta^{}_W}}

\def\acF{{\delta \eta^{}_F\delta \phi^{}_F/2\pi}}
\def\acB{{\delta \eta^{}_B\delta \phi^{}_B/2\pi}}
\def\acW{{\delta \eta\delta \phi/2\pi}}

\def\eg{{\eta^{}_{gap}}}
\def\yg{{\eta^{}_{gap}}}
\def\fg{{\phi^{}_{gap}}}

\def\omN{{\omega_N}}

\def\bp{\textbf{p}}
\def\oq{\overline{q}}
\def\fv{\phi}
\def\loy{\lambda_1(\eta)}
\def\loyo{\lambda_1(\eta_1)}
\def\loyt{\lambda_1(\eta_2)}
\def\lty{\lambda_2(\eta_1,\eta_2)}
\def\lo#1{\lambda_1(#1)}
\def\loo#1{\lambda^2_1(#1)}
\def\lt#1{\lambda_2(#1)}
\def\Lam#1{\Lambda(#1)}
\def\tLam#1{\widetilde{\Lambda}(#1)}

\def\loyf{\lambda_1(\eta,\phi)}
\def\loyfo{\lambda_1(\eta_1,\phi_1)}
\def\loyft{\lambda_1(\eta_2,\phi_2)}
\def\ltyf{\lambda_2(\eta_1,\phi_1;\eta_2,\phi_2)}

\def\roy{\rho_1(\eta)}
\def\royo{\rho_1(\eta_1)}
\def\royt{\rho_1(\eta_2)}
\def\rty{\rho_2(\eta_1,\eta_2)}
\def\ro#1{\rho_1(#1)}
\def\roo#1{\rho^2_1(#1)}
\def\rt#1{\rho_2(#1)}
\def\royf{\rho_1(\eta,\phi)}
\def\royfo{\rho_1(\eta_1,\phi_1)}
\def\royft{\rho_1(\eta_2,\phi_2)}
\def\rtyf{\rho_2(\eta_1,\phi_1;\eta_2,\phi_2)}

\def\IFF{{I_{F\!F}^{}}}
\def\IBF{{I_{F\!B}^{}}}
\def\IBB{{I_{B\!B}^{}}}
\def\JFF{{J_{F\!F}^{}}}
\def\JBF{{J_{F\!B}^{}}}
\def\JFB{{J_{F\!B}^{}}}
%%%%%%%%%%%%%%%%%%%%%

In the past few decades, considerable attention has been devoted to the experimental \cite{Uhlig78}-\cite{ATLAS12} and theoretical \cite{CapKr78}-\cite{OlszBron13} exploration of ``forward-backward" (FB) correlations in high-energy pp and AA collisions.
This refers to correlation between the multiplicities of charged particles produced in forward, $\nF$, and backward, $\nB$, separated rapidity windows.
One of the challenges in these investigations is the isolation of the ``volume" contribution, which originates from an event-by-event fluctuation in the number of emitting sources \cite{CapKr78}.

It has been previously suggested in \cite{LMcL10} to use the event multiplicity in a third rapidity window
to solve this problem, but as discussed in \cite{Bzdak12}, this complicates the interpretation of results.
In the present paper we argue that by studding the FB
multiplicity correlation between windows separated in both rapidity and azimuth,
the volume contribution can be isolated.
Additional important quantitative physical information about the magnitude of this event-by-event fluctuation that causes this contribution can also be obtained.

We also show that the traditional definition of the FB correlation coefficient has a strong dependence
on the acceptance size of the windows, which causes the coefficient to go to zero with the acceptance.
Consequently, results obtained from windows of different widths cannot be compared directly.
In this paper, we propose suitable observables for FB correlation studies, which have finite, nonzero limit
as the window acceptances go to zero.

To check our observations we use the simple two stage model \cite{PLB00,EPJC04,Vestn1}, inspired by
a string picture of hadronic interactions.
This model suggests that during the initial stage of the interaction, some number
of strings are formed, which are then considered as identical, independent emitters of charged particles.
In our note \cite{Dub10} we considered only the long-range (LR) part of the correlation, originating
from the fluctuation in the number of sources (the strings or as suggested
in \cite{CapKr78} the cut pomerons).
In the present paper we also
take into account
the short-range (SR) correlation
between particles produced by a single string.
This SR correlation can arise from several distinct physical processes such as
the formation and decay of clusters, resonances, or minijets during the string fragmentation.
We note that the presence of such SR correlations, along with the influence
on the FB multiplicity correlation, inevitably turns a string into non-poissonian emitter.

We show that studying the FB multiplicity correlation between windows separated both in rapidity and azimuth
allows for the separation of the LR and SR contributions.
We also demonstrate using a model independent analysis that determining the FB multiplicity correlation coefficient between two small windows separated in rapidity and azimuth also determines the two-particle correlation function $C_2$.
This even holds if the particle distribution in rapidity is not flat (as e.g. in the case of pA interactions) and the $C_2$ does not only depend on the differences of rapidities.

The paper is organized as follows. In Sec.~2 we discuss the different versions of the definition of the FB correlation coefficient and generalize this definition for the case of windows separated both in rapidity and azimuth.
In Sec.~3 we outline the connection between FB correlation coefficient and the two-particle correlation function $C_2$.

In Sec.~4 we
formulate a two stage model
with strings as independent identical sources, introduce the pair correlation function of a single string, and calculate  the FB correlation coefficient.  This includes describing the LR and SR contributions in the framework of this model.
In Sec.~5 we parameterize the pair correlation function of a single source in accordance with
the string decay picture.
We then fit the parameters using the data on
the FB correlation strength
between multiplicities in small azimuth and rapidity windows.

In Sec.~6 we use the resulting model
to calculate the values of the FB correlation coefficient for large rapidity windows
of different width and separation
and compare the results with the preliminary experimental data from ALICE \cite{PoS12Feof}.
In Sec.~7 we introduce suitable alternative observables for future FB multiplicity correlation studies.

In Appendix A we describe the calculation of integrals over rapidity and azimuth windows.
In Appendix B we present an alternative derivation of the basic formulae (\ref{C2_Lam}) and (\ref{blarg})
and check that the resulting expression for the FB correlation coefficient is consistent with the expression for the LR correlation coefficient obtained earlier in \cite{Dub10} in the large rapidity separation limit.
Finally, in Appendix C we discuss the correspondence between the FB multiplicity correlations in windows separated in azimuth and rapidity and the ``untriggered di-hadron" correlations.

\section{Definition of the FB Correlation Coefficient} \label{sec:def}
The FB correlation coefficient is traditionally \cite{Uhlig78,UA5_83,E735,STAR09} defined as the coefficient, $b$, used  in linear regression:
\beq \label{corf}
\avr\nB\nF =a+b\, \nF  \ .
\eeq
In this case
\beq \label{defb}
b =\frac{\av{\nF\nB}-\av{\nF}\av{\nB}} {D_\nF} \ ,
\eeq
where $D_\nF$ is the variance of the multiplicity in the forward window:
\beq \label{DnF}
D_\nF =\av{n_F^2}-\av{\nF}^2_{} \ .
\eeq	

Clearly, this value of such a coefficient would depend on
the acceptances of the forward and/or backward windows.
To avoid this trivial influence one can shift from $\nF$ and $n_B$ to the relative
or scaled observables \cite{PPR} $\nu_F^{}=\nF/\av\nF$ and $\nu_B^{}=\nB/\av\nB$.
In these observables $\avr{\nu_B^{}}{\nu_F^{}} =a_{rel}^{}+\brel\, {\nu_F^{}}$ and
\beq \label{brel}
\brel =\frac{\av{\nu_F^{}\nu_B^{}}-1} {\av{\nu_F^{2}}-1}=\frac{\av\nF}{\av\nB}\, b \ ,
\eeq

The following symmetrized form of (\ref{defb}) is also used \cite{UA5_88,ATLAS12}
\beq \label{defbsym}
\bsym =\frac{\av{\nF\nB}-\av{\nF}\av{\nB}} {\sqrt{D_\nF D_\nB}}  \ ,
\eeq
where it can be proven that $|\bsym|\leq 1$. Note that for symmetric window where $\av\nF=\av\nB$ and $D_\nF=D_\nB$, all these definitions converge to  the same result:
\beq \label{bequal}
\brel =\bsym =b
\ .
\eeq

In present paper we study the multiplicity correlations between $\nF$ and $n_B$ in windows separated both in rapidity and in azimuth. We denote the width of the forward and backward windows in rapidity and in azimuth using  $\DyF$, $\DfF$, and $\DyB$, $\DfB$, respectively.
Likewise, we denote the positions of the centers of these window using $\yF$, $\fF$ and $\yB$, $\fB$, respectively.
We will also introduce the following short notation for the acceptance of  forward and backward windows:
\beq \label{ac}
\aF\equiv \acF  \ , \hs1  \aB\equiv \acB  \ .
\eeq
Finally, we denote the distances between the centers of these windows as follows:
\beq \label{cc}
\yFB\equiv \yF-\yB  \ , \hs1  \fFB\equiv\fF-\fB
\eeq

These variables are trivially related to the rapidity gap, $\eta^{}_{gap}$, and azimuthal gap, $\fv^{}_{gap}$,
between the two windows:
\beq \label{gap}
\yFB=\frac{\DyF}{2}+\eta^{}_{gap}+\frac{\DyB}{2}  \ , \hs1  \fFB=\frac{\DfF}{2}+\fv^{}_{gap}+\frac{\DfB}{2} \ ,
\eeq
For symmetric windows, where $\DyF=\DyB=\Dy$ and $\DfF=\DfB=\Df$, this simplifies further:
\beq \label{gapsym}
\yFB=\eta^{}_{gap}+\Dy  \ , \hs1  \fFB=\fv^{}_{gap}+\Df    \ .
\eeq

\section{Connection with Two-Particle Correlation Function} \label{twopart} One can express the FB correlation coefficient through the two-particle correlation function
 $C_2(\eta_1,\eta_2;\phi_1,\phi_2)$.
First, we introduce the one, $\royf$, and two-particle, $\rtyf$, charge particle densities:
\beq \label{den}
\royf=\frac{d^2N}{d\eta \,d\fv}
\ , \hs1
\rtyf=\frac{d^4N}{d\eta _1\,d\fv_1\ d\eta _2\,d\fv_2}
 \ .
\eeq
By integrating (\ref{den}) over the forward acceptance interval, $\eta\!\in\!\DyF$, $\fv\!\in\!\DfF$, we obtain \cite{Voloshin02}:
\beq \label{rofF}
\int_\DyfF \!\!\dyf\, \royf = \av{\nF} \ ,
\eeq
$$
\int_\DyfF \!\!\dyfo \int_\DyfF \!\!\dyft\, \rtyf =\av{\nF(\nF-1)}  \ .
$$
When we integrate over the forward, $\eta_1\!\in\!\DyF$, $\fv_1\!\in\!\DfF$,
and the backward, $\eta_2\!\in\!\DyB$, $\fv_2\!\in\!\DfB$, acceptance intervals, we have
\beq \label{rofBF}
 \int_\DyfF \!\!\dyfo \int_\DyfB  \!\!\dyft\, \rtyf =\av{\nF\nB} \ .
\eeq
Here, $ \av{\nF}$ is the average multiplicity produced in the acceptance $\DyfF$.

For windows of sufficiently small rapidity and azimuthal acceptance, (\ref{rofF}) and (\ref{rofBF}) imply:
\beq \label{roFB-ex}
\rho_1^{}(\yF,\fF) = \frac{\av{\nF}}{\DyfF} \ ,   \hs 1
\rho_2^{}(\yF,\fF;\yB,\fB) = \frac{\av{\nF\nB}}{\DyfF\DyfB} \ ,
\eeq
\beq \label{roF-ex}
\rho_2^{}(\yF,\fF;\yF,\fF) = \frac{\av{\nF(\nF-1)}}{(\DyfF)^2_{}} \ .
\eeq
These formulae allow for the experimental measurement of $\royf$ and \\ $\rtyf$, and hence the two-particle correlation function $C_2$.  We introduce this function in the standard way:
\beq \label{C_2}
C_2(\eta_1,\eta_2;\phi_1,\phi_2)=\frac{\rt{\eta _1,\eta_2;\phi_1,\phi_2}}{\rho_1^{}(\eta_1,\phi_1)\rho_1^{}(\eta_1,\phi_2)}-1 \ .
\eeq
Substituting  (\ref{roFB-ex}) into (\ref{C_2}), we obtain that for windows of small rapidity and azimuthal acceptance:
\beq \label{C2ex}
C_2(\yF,\fF;\yB,\fB)=
\frac{\av{\nF\nB}-\av{\nF}\av{\nB}}{\av{\nF}\av{\nB}}
= \avL{\nFr\nBr}-1    \ .
\eeq

It is important to note that by (\ref{C2ex}), the measurement of multiplicity correlation between two sufficiently small window determines the two-particle correlation function $C_2$ in accordance with the standard definition in (\ref{C_2}).
This also holds in the absence of the translational invariance in rapidity. It does not necessitate an event mixing procedure, which is usually applied in the di-hadron correlation analysis, as discussed in Appendix C.

Note that if we change  $\nF$ and $\nB$ to denote the multiplicities of particles with  transverse momenta in
intervals $\dpF$ and $\dpB$, respectively, then (\ref{C2ex}) can be used
to measure the two-particle correlation function $C_2$ between particles belonging to these intervals.
This is analogous to what is done in the triggered di-hadron correlation approach.
In principle, this enables using the small $\dpF$ and $\dpB$ intervals to measure the whole two-particle correlation function $C_2(\bpF,\bpB)$, where the 3-momenta  $\bpF$ and $\bpB$ are the centers of the $\dyF\,\dfF\,\dpF$ and  $\dyB\,\dfB\,\dpB$ intervals.

The following simplifications can be made due to azimuth rotation invariance:
\beq \label{ro_rot}
\ro{\eta ,\phi}=\roy/2\pi   \ , \hs1
\rtyf =\rt{\eta _1,\eta_2;\phi_1-\phi_2} /(2\pi)^2_{}
\eeq
\beq \label{C_2_1}
C_2(\eta_1,\eta_2;\phi_1-\phi_2)=\frac{\rt{\eta _1,\eta_2;\phi_1-\phi_2}}{\royo\royt}-1 \ .
\eeq

For windows of arbitrary widths, (\ref{rofF}) and (\ref{rofBF}) imply:
\beq \label{nBF}
\av{\nF\nB}-\av{\nF}\av{\nB}=\av{\nF}\av{\nB}\IBF \ ,
\eeq
\beq \label{D_int-r}
D_\nF  = \av{\nF} + \av{\nF}^2_{}\IFF   \ ,
\eeq
where
\beq \label{rofF1}
\av{\nF}=\frac{\DfF}{2\pi}\int_\DyF \!\!d\eta\, \roy   \ ,
\eeq
\beq \label{IBF1}
 \IBF=\frac{1}{(2\pi)^2_{}\av{\nF}\av{\nB}}\int_\DyfF \!\!\!\!\!\!\!\!\dyfo \int_\DyfB \!\!\!\!\!\!\!\!\dyft\,  \royo\royt
 C_2(\eta_1,\eta_2;\phi_1-\phi_2)   ,
\eeq
\beq \label{IFF1}
\IFF=\frac{1}{(2\pi)^2_{}\av{\nF}^2_{}}\int_\DyfF \!\!\!\!\dyfo \int_\DyfF \!\!\!\!\dyft\, \royo\royt
C_2(\eta_1,\eta_2;\phi_1-\phi_2)  .
\eeq
This gives the following expression for the correlation coefficient:
\beq \label{bet_int-r}
 \brel=\frac{\av\nF}{\av\nB}\, b   =\frac{\av{\nF}\IBF}{1+\av{\nF} \IFF}     \ .
\eeq
Note that in the absence of correlation, i.e. when
$C_2=0$,  $\IBF=\IFF=0$ by definition which means that $D_\nF  = \av{\nF}$ by  (\ref{D_int-r}).
 For further simplifications of the integrals (\ref{IBF1}) and (\ref{IFF1}), see Appendix A.

In the appendix we show that for the case of FB windows are only separated in rapidity, i.e. when $\DfF=\DfB=2\pi$, (\ref{Int_2pi}) gives:
\beq \label{rofF2}
\av{\nF}=\int_\DyF \!\!d\eta\, \roy   \ ,
\eeq
\beq \label{IBF}
 \IBF=\frac{1}{\av{\nF}\av{\nB}}\int_\DyB \!\!d\eta_1 \int_\DyF \!\!d\eta_2\,  \royo\royt C_2(\eta_1,\eta_2)  \ ,
\eeq
\beq \label{IFF}
\IFF=\frac{1}{\av{\nF}^2_{}}\int_\DyF \!\!d\eta_1 \int_\DyF \!\!d\eta_2\, \royo\royt C_2(\eta_1,\eta_2)  \ ,
\eeq
where
\beq \label{C2}
C_2(\eta_1,\eta_2) =
\frac{1}{\pi}\int_{0}^{\pi} \!\!d\fv \,  C_2(\eta_1,\eta_2;\fv)  \ .
\eeq

For sufficiently small windows in rapidity and azimuth, in which  $C_2(\eta_1,\eta_2;\phi_1-\phi_2)$ and $\ro{\eta }$ can be treated as constant, we have:
\beq \label{smallmF-r}
 \av{\nF}= \ro{\yF} \aF  \ , \hs1   \av{\nB}= \ro{\yB} \aB  \ ,
\eeq
\beq \label{IBFsm-r}
 \IBF= C_2(\yF,\yB;\fFB) \ ,
\eeq
\beq \label{IFFsm-r}
\IFF=
C_2(\yF,\yF;0)  \ ,
\eeq
\beq \label{D_sm-r}
D_\nF
= \av{\nF}[1 +
\av{\nF} C_2(\yF,\yF;0)]  \ ,
\eeq
\beq \label{bet_sm-r}
 \brel=\frac{\av\nF}{\av\nB}\, b  =\frac{\av{\nF}  C_2(\yF,\yB;\fFB)}{1 +
\av{\nF} C_2(\yF,\yF;0)}   \ .
\eeq
Recalling our short notations (\ref{ac}) and (\ref{cc}), we see that the correlation coefficient (\ref{brel}),
even defined in scaled variables, still depends through $\av\nF$ on the acceptance $\aF$ of the forward window.
This was observed earlier \cite{Vestn1,Dub10} in the framework of a simple model.

When both of the sufficiently small FB windows are situated in the central region,
the translational invariance in rapidity holds:
\beq \label{C_2_2}
\ro{\eta }=\rhoo  \ , \hs{0.3}
 \rho^{}_2(\eta_1,\eta_2;\fv)= \rho^{}_2(\eta_1-\eta_2;\fv)  \ , \hs{0.3}
 C_2(\eta_1,\eta_2;\phi)= C_2(\eta_1-\eta_2;\phi).
\eeq
The  formulae (\ref{smallmF-r})--(\ref{bet_sm-r}) can then be further simplified:
\beq \label{smallmF-r1}
 \av{\nF}= \rhoo \aF  \ , \hs1   \av{\nB}= \rhoo \aB  \ ,
\eeq
\beq \label{D_sm-r1}
D_\nF
= \av{\nF}[1 +
\aF\rhoo C_2(0,0)]  \ ,
\eeq
\beq \label{br_sm1}
\brel=\frac{\aF}{\aB}\, b
=\frac{\aF\rhoo C_2(\yFB,\fFB)}{1+\aF\rhoo C_2(0,0)}  \ .
\eeq

Assuming  (\ref{C_2_2}) and (\ref{smallmF-r1}), the FB correlation between large windows situated
in the central rapidity region
can be fully described using the formulae (\ref{D_int-r}) and (\ref{bet_int-r}),
with the following relevant expressions for $\IBF$ and $\IFF$:
\beq \label{IBFsm}
 \IBF=  (\DyfF\DyfB)_{}^{-1}\int_\DyfF \!\!\dyfo \int_\DyfB \!\!\dyft\,   C_2(\eta_1-\eta_2;\phi_1-\phi_2)
 \ ,
\eeq
\beq \label{IFFsm}
\IFF= (\DyfF)_{}^{-2}\int_\DyfF \!\!\dyfo \int_\DyfF \!\!\dyft\,  C_2(\eta_1-\eta_2;\phi_1-\phi_2)
\eeq
For further simplification of these integrals, see Appendix A.

\section{The Model} \label{model} In this section, we calculate the FB correlations between windows separated in rapidity and azimuth using the simple two stage model \cite{PLB00,EPJC04,Vestn1}.  This model inspired by the string picture of hadronic interactions.
In this model we suggest that at the initial stage of interaction, $N$ strings are formed.
This value fluctuates event-by-event resulting in the scaled variance:
\beq \label{omN}
\omega_N=D_N/\av N=(\av {N^2}-\av {N}^2)/\av N \ .
\eeq
Note that this fluctuation is non-poisson in both pp and AA collisions \cite{PRC11}, hence $\omega_N\neq 1$.
Furthermore, its value depends on the collision energy.
After the initial stage of the interaction, one considers these strings
as identical independent sources of observed charge particles.

Along with the long-range (LR)
part of the correlation \cite{CapKr78,EPJC04, Dub10}, originating the fluctuation in $N$,
this section also takes into account the short-range (SR) part.
The SR part originates from the correlation between particles produced by a single string.

\subsection{Pair Correlation Function of a Single String}To characterize the correlation of particles produced from a string,
we proceed in analogy to the consideration in the Sec.~ \ref{twopart}.
We start by introducing the two-particle correlation function for charged particles
produced from the decay of a single string:
\beq \label{Lam1}
\Lambda(\eta_1,\eta_2;\phi_1\!-\!\phi_2)=\frac{\lt{\eta_1,\eta_2;\phi_1\!-\!\phi_2}}{\loyo\loyt}-1 \ .
\eeq
In the above equation,  $\loy$ and $\lt{\eta_1,\eta_2;\phi_1\!-\!\phi_2}$ respectively denote
the one and two-particle densities of charge particles produced from one string.
We assume that the particle emission from a string is isotropic in the azimuthal angle, $\phi$.

Similarly to (\ref{rofF}) and (\ref{rofBF}) we note that
\beq \label{lamfF}
\int_\DyfF \!\!\dyf\, \loyf =\omF \ ,
\eeq
$$
\int_\DyfF \!\!\dyfo \int_\DyfF \!\!\dyft\, \lt{\eta_1,\eta_2;\phi_1\!-\!\phi_2} =\av{\mF(\mF-1)}    \ ,
$$
\beq \label{lamfBF}
 \int_\DyfF \!\!\dyfo \int_\DyfB  \!\!\dyft\, \lt{\eta_1,\eta_2;\phi_1\!-\!\phi_2} =\av{\mB\mF}    \ ,
\eeq
where $\omF$ is the average multiplicity produced by one string in the forward window, $\DyfF$, and $\omB$ is the corresponding variable in the backward window, $\DyfB$.

Using (\ref{Lam1})--(\ref{lamfBF}), we can write:
\beq \label{mBF}
\av{\mB\mF}-\av{\mB}\av{\mF}=\av{\mB}\av{\mF}\JBF \ ,
\eeq
\beq \label{D_mF}
D_\mF  = \av{\mF} + \av{\mF}^2_{}\JFF   \ ,
\eeq
where
\beq \label{lamfF1}
 \omF=\frac{\DfF}{2\pi}\int_\DyF \!\!d\eta\, \loy  \ ,
\eeq
\beq \label{JBF}
 \JBF=\frac{1}{(2\pi)^2_{}\av{\mF}\av{\mB}}\int_\DyfF \!\!\!\!\!\!\!\!\dyfo \int_\DyfB \!\!\!\!\!\!\!\dyft\,  \loyo\loyt
 \Lam{\eta _1,\eta_2;\phi_1-\phi_2}    ,
\eeq
\beq \label{JFF}
\JFF=\frac{1}{(2\pi)^2_{}\av{\mF}^2_{}}\int_\DyfF \!\!\!\!\!\!\dyfo \int_\DyfF \!\!\!\!\!\!\dyft\, \loyo\loyt
\Lam{\eta _1,\eta_2;\phi_1-\phi_2}    .
\eeq
We note that (\ref{D_mF}) implies that presence of SR correlation turns the string into non-poissonian emitter.

For sufficiently small windows, in which we can
consider $\Lam{\eta _1,\eta_2;\phi_1-\phi_2}$ and $\lo{\eta }$  to be constant,
we find:
\beq \label{smallmF}
\omF= \lo{\yF} \aF  \ , \hs1  \omB= \lo{\yB} \aB  \ ,
\eeq
\beq \label{JBFsm}
 \JBF=  \Lam{\yF,\yB;\fFB} \ ,
\eeq
\beq \label{JFFsm}
\JFF=  \Lam{\yF,\yF;0}  \ ,
\eeq
where we have used our notations for $\aF$ and $\aB$ as in (\ref{cc}) for the window acceptances.

If both of the small windows are situated in the central rapidity region, where each string contributes
to the particle production in the selected rapidity range, then the translational invariance in rapidity implies:
\beq \label{Lam2}
\lo{\eta }=\mo    \ , \hs1
\Lam{\eta _1,\eta_2;\phi}=\Lam{\eta _1-\eta_2;\phi}
\eeq
and the  formulae  (\ref{smallmF})--(\ref{JFFsm})  take the form
\beq \label{smallmF1}
\omF= \mo \aF  \ , \hs1  \omB= \mo \aB  \ ,
\eeq
\beq \label{JFBsm1}
\JBF=\Lam{\yFB,\fFB} \ ,
\eeq
\beq \label{JFFsm1}
\JFF=\Lam{0,0}  \ .
\eeq
Recall that $\yFB$ and $\fFB$ are the distances between the centers of forward
and backward windows in rapidity and azimuth, as denoted in (\ref{cc}).

Assuming (\ref{smallmF1}), large windows situated in the central rapidity region
can also be fully described using the formulae (\ref{mBF}) and (\ref{D_mF}),
noting the
following relevant expressions for $\JBF$ and $\JFF$:
\beq \label{subst_int1}
\JBF=(\DyfF\DyfB)_{}^{-1} \int_\DyfF \!\!\dyfo \int_\DyfB \!\!\dyft\,  \Lam{\eta_1-\eta_2,\phi_1-\phi_2} \ ,
\eeq
\beq \label{subst_int2}
\JFF=(\DyfF)_{}^{-2} \int_\DyfF \!\!\dyfo \int_\DyfF \!\!\dyft\, \Lam{\eta_1-\eta_2,\phi_1-\phi_2}  \ .
\eeq
See Appendix A for further simplification of these integrals.

\subsection{Resulting Correlation Strength} \label{correl} In considered model with strings as independent identical sources (SM)
 one can write for $N$ sources
\cite{Voloshin02}:
\beq \label{roNo}
\rho^{N}_1(\eta)=N  \lambda_1(\eta) \ ,
\eeq
\beq \label{roNt}
\rho^{N}_2(\eta_1,\eta_2;\fv)=N \lambda_2(\eta_1,\eta_2;\fv) + N(N-1)\lambda_1(\eta_1) \lambda_1(\eta_2)   \ .
\eeq
The relevant charged particle densities in (\ref{den}) are then given by
\beq \label{ro1-av}
\rho^{}_1(\eta)=\av{\rho^{N}_1(\eta)}=\av{N}\lambda_1(\eta_1) \ ,
\eeq
$$
\rho^{}_2(\eta_1,\eta_2;\fv)=\av{\rho^{N}_2(\eta_1,\eta_2;\fv)}=
$$
\beq \label{ro2-av}
=\av N [\lambda_2(\eta_1,\eta_2;\fv) -\lambda_1(\eta_1) \lambda_1(\eta_2)]
+ \av{N_{}^2}\lambda_1(\eta_1) \lambda_1(\eta_2)   \ .
\eeq
This leads to the following connection between correlators:
$$
\rho^{}_2(\eta_1,\eta_2;\fv)-\rho^{}_1(\eta_1)\rho^{}_1(\eta_2)=
$$
\beq \label{cor-av}
=\av N [(\lambda_2(\eta_1,\eta_2;\fv) -\lambda_1(\eta_1) \lambda_1(\eta_2)]+ D_N^{} \lambda_1(\eta_1) \lambda_1(\eta_2)   \ ,
\eeq
where $D_{N}$ is the event-by-event variance in the number of sources, (\ref{omN}).
This results in the
following expression for the two-particle correlation function $C_2(\yF,\yB;\fFB)$ (\ref{C_2_1}):
\beq \label{C2_Lam}
C_2(\eta_1,\eta_2;\fv)=\frac{\omega_N+ \Lam{\eta _1,\eta_2;\fv}}{\av N}
\ ,
\eeq
where $\omega_N$ is the scaled variance of the number of sources, (\ref{omN}), and $\Lam{\yF,\yB;\fFB}$ is
the pair correlation function of a single string, (\ref{Lam1}).

We note that constant term in (\ref{C2_Lam}) is physically important.
Its magnitude, $\omega_N/\av N = D_N/\av {N}^2$, corresponds to the magnitude of the fluctuation
of the number of  sources \cite{CapKr78}.
The last depends on the initial energy and the fixation of collision centrality.

For sufficiently small windows in both rapidity and azimuth, the expression for $C_2$ in (\ref{C2_Lam}) can be used in (\ref{bet_sm-r}) to recover the expression for the FB correlation coefficient (\ref{brel}):
\beq \label{bsm-y}
\brel =\frac{\av\nF}{\av\nB}\, b
=\frac{\av\nF[\omega_N+ \Lam{\yF,\yB;\fFB}]/\av N}{1+\av\nF[\omega_N+ \Lam{\yF,\yF;0}]/\av N}
\ .
\eeq

If these small FB windows are situated in the central rapidity region, where translational invariance in rapidity holds, then $\Lam{\yF,\yB;\fFB}$ will only depend on the difference of rapidities, $\yFB=\yF-\yB$.  Hence, (\ref{bsm-y}) can be simplified to the following form:
\beq \label{bsm}
\brel =\frac{\aF}{\aB}\, b
=\frac{\aF\mo[\omega_N+ \Lam{\yFB,\fFB}]}{1+\aF\mo[\omega_N+ \Lam{0,0}]}
\
\eeq
where $\mo$ is the average rapidity density of charged particles produced by one string.
See Appendix B for an alternative derivation of
(\ref{C2_Lam}) for this case.

The resulting FB correlation coefficient in (\ref{bsm}) can be described as the sum of two terms, $\brel=b_{rel}^{LR}+b_{rel}^{SR}$, where:
\beq \label{bLsm}
b_{rel}^{LR}=\frac{\aF\mo\omega_N}{1+\aF\mo[\omega_N+ \Lam{0,0}]}
 \ ,
\eeq
and
\beq \label{bSsm}
b_{rel}^{SR}=\frac{\aF\mo}{1+\aF\mo[\omega_N+ \Lam{0,0}]}\Lam{\yFB,\fFB}
 \ .
\eeq
The first term is dependent on the acceptance $\aF$ of the forward window, but independent of the distances in rapidity, $\yFB$, and azimuth, $\fFB$, between the forward and backward windows.
This is why this term is refereed to as the long range (LR) contribution.
This contribution manifests as a constant pedestal in the plot of the FB correlation coefficient $b$  against $\yBF$
and $\fBF$, Figs.~\ref{sm9} and \ref{sm7} below.
The height of this pedestal is determined by the event-by-event fluctuation of the number of the strings (sources) $N$
and can be used to evaluate of the extent of this fluctuation.
Note that at any fixed number of sources there will be no such contribution, as $\omega_N\equiv D_N/\av{N}=0$.

The second term is proportional to the pair correlation function $\Lam{\yFB,\fFB}$
of a single string, which is scaled by a common factor that depends on the acceptance $\aF$
of the forward window (\ref{ac}).
This contribution gives structure to the FB correlation coefficient $b$ above the constant term
when plotted against $\yBF$ and $\fBF$.  Hence, it is referred to as the short range (SR) contribution.

We would like to emphasize that if the pair correlation function of a single string is equal to zero, $\Lam{\yFB,\fFB}=0$, such that there are no SR correlations, $b_{\Lambda=0}^{SR} =0$, we still have nonzero FB correlations due to the LR contribution:
\beq \label{bsm1}
b_{rel}^{\Lambda=0}=b_{\Lambda=0}^{LR} =\frac{\aF\mo\omega_N}{1+\aF\mo\omega_N}
\ ,
\eeq
which originates from the event-by-event fluctuation in the number of strings, $N$.
We also note
that for $\Lambda=0$, (\ref{D_mF}) and (\ref{JFF}) imply that the multiplicity distribution
from a string becomes poissonian, $D_\mF  = \av{\mF}$,
and the expression for the correlation coefficient in (\ref{bsm1}) agrees
with the results obtained in \cite{EPJC04,Vestn1,Dub10} for this case.

For windows of an arbitrary width in rapidity and azimuth in which one cannot take $\Lam{\eta _1,\eta_2;\phi}$ to be constant, equations (\ref{nBF})--(\ref{bet_int-r}) and (\ref{C2_Lam}) can be used to rewrite (\ref{bsm-y}) as:
\beq \label{blarg}
\brel =\frac{\av\nF}{\av\nB}\, b
=\frac{\av\mF[\omega_N+ \JFB]}{1+\av\mF[\omega_N+ \JFF]}    \ ,
\eeq
where $\omF$ is the mean multiplicity produced in the forward window by a single string as in (\ref{lamfF1}), and the integrals $\JFB$ and $\JFF$ are given by (\ref{JBF}) and (\ref{JFF}).
If both of these windows are in the central rapidity region where translational invariance holds,  integrals $\JFB$ and $\JFF$ can be simplified
to (\ref{subst_int1}) and (\ref{subst_int2}).

For the noteworthy case of symmetric FB windows ($\DyF=\DyF=\Dy$),
separated only in rapidity, i.e. with full azimuthal acceptance
  $\DfF=\DfB=2\pi$,
the expressions for $\JFB$ and $\JFF$ in (\ref{subst_int1}) and (\ref{subst_int2}) can be reexpressed:
\beq \label{subst_int3}
\JFB=\frac{1}{\Dy^2}  \int_{-\Dy}^\Dy \!\!d\eta\,  \Lam{\eta +\yFB} \,t_\Dy(\eta)
  \ ,
\eeq
\beq \label{subst_int4}
\JFF=\frac{2}{\Dy^2} \int_{0}^\Dy \!\!d\eta\,  \Lam{\eta }  (\Dy-\eta)
= \frac{2}{\Dy} \int_{0}^\Dy  \!\!\Lam{\eta } \,d\eta  -\frac{2}{\Dy^2} \int_{0}^\Dy   \!\!\Lam{\eta } \,\eta\,d\eta
  \ ,
\eeq
where
\beq \label{Lam-y}
\Lam{\eta }= \frac{1}{\pi}\int_{0}^{\pi} \!\!d\fv \,  \Lam{\eta ,\fv}
\eeq
and
$t_\Dy(\eta)$ is the "triangular" weight function, (\ref{tri}).  This weight arises at the integration
of (\ref{subst_int1}) and (\ref{subst_int2}) over $(\eta_1\!+\!\eta_2)/2$
and reflects the corresponding phase space. See Appendix A for more details.

\section{Rapidity-Azimuth Dependence of the FB Multiplicity Correlation} \label{topol}
In this section, we are fixing the model parameters using the data on the
dependence of the FB correlation coefficient
on the rapidity and azimuthal distances, $\yFB$ and $\fFB$, between the centers of two
small FB windows.

\subsection{Parametrization of the Pair Correlation Function of a Single String}As shown in Sec.~\ref{model}, the pair correlation function of string $ \Lam{\eta,\phi}$
is needed to calculate the short-range contribution to the FB multiplicity correlation strength.

\begin{figure}[t]
\centerline{
\includegraphics[width=80mm,angle=0]{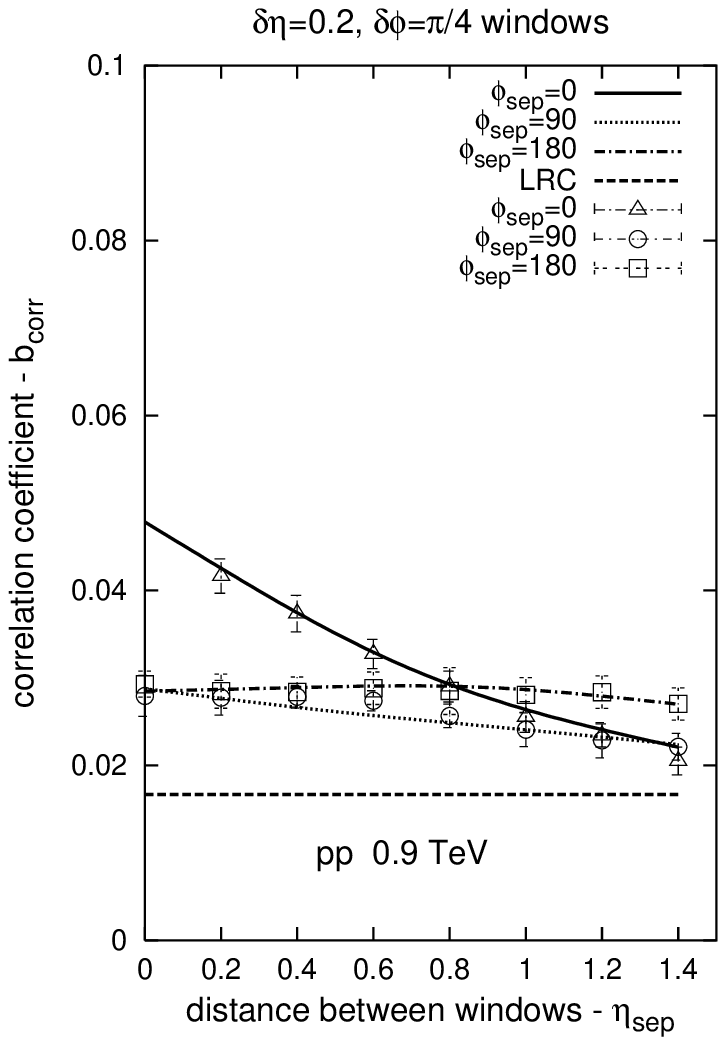}
\includegraphics[width=80mm,angle=0]{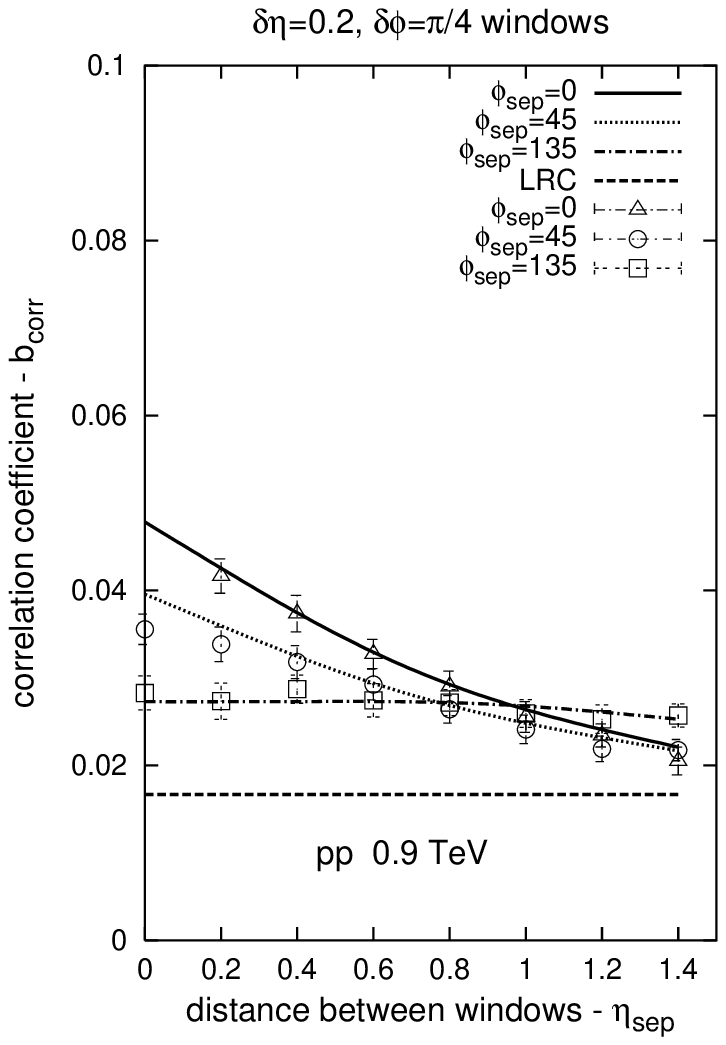}
}
\caption{\label{sm9}
The forward-backward (FB) correlation coefficient, (\ref{defb}), in pp collisions at 0.9 TeV for small, symmetric windows with $\DyF=\DyB=\Dy=0.2$ rapidity acceptance and $\DfF=\DfB=\Df=\pi/4$ azimuthal acceptance.
These values are calculated by accounting for both the long-range (LR), (\ref{bLsm}), and short-range (SR),
(\ref{bSsm}), contributions, as described in (\ref{blarg}),  (\ref{sub_int1}) and (\ref{sub_int2}).
Its value is plotted as a function of distance between the centers of windows in rapidity $\yFB=\eta_{sep}$, and compared at various distances between the centers of windows in azimuth, $\fFB=\phi_{sep}$.
The left panel compares $\phi_{sep}=0^\circ$, $90^\circ$, $180^\circ$  azimuthal separations, while the right panel compares $\phi_{sep}=0^\circ$, $45^\circ$, $135^\circ$ .
The corresponding experimental data points are taken from \cite{AltsPhD}.
}
\end{figure}
\begin{figure}[t]
\centerline{
\includegraphics[width=80mm,angle=0]{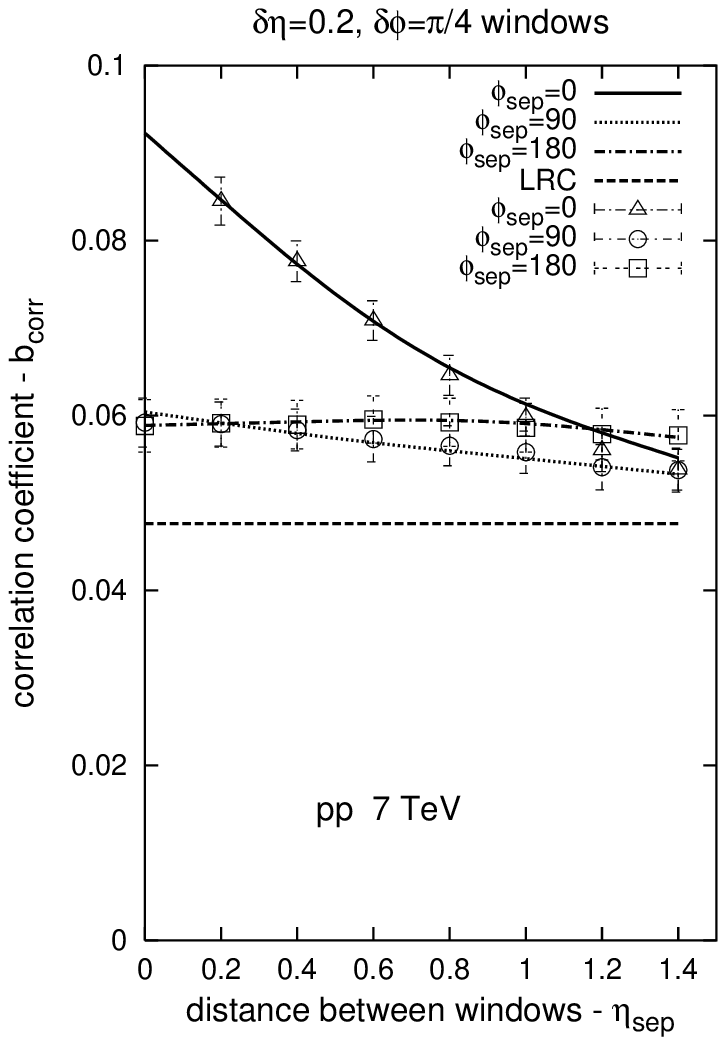}
\includegraphics[width=80mm,angle=0]{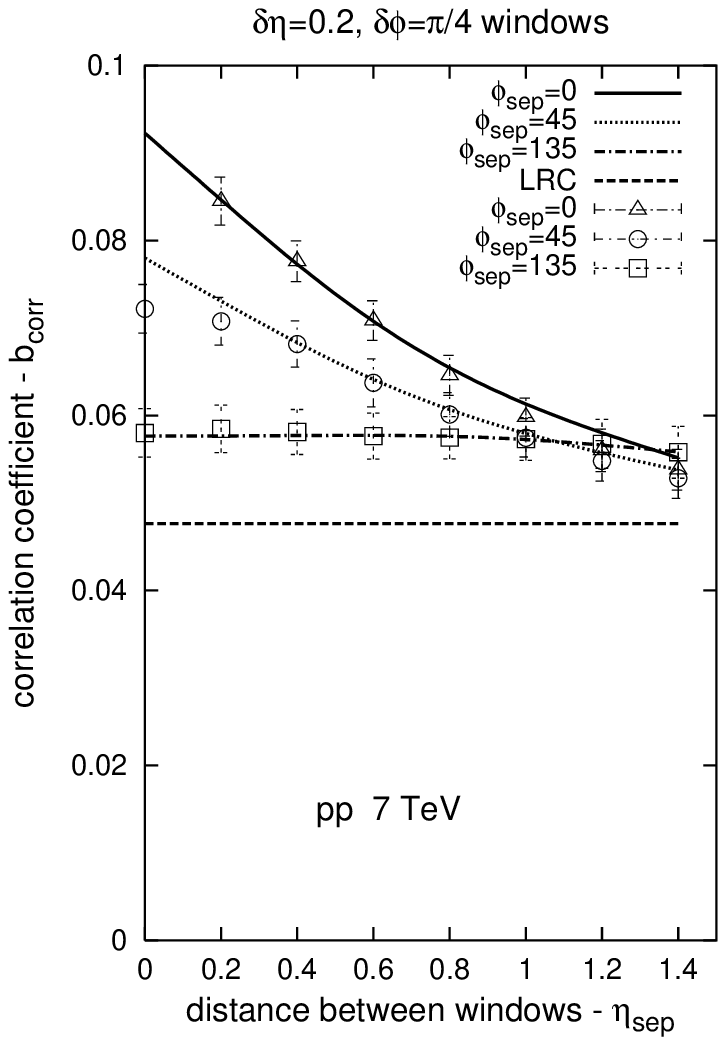}
}
\caption{\label{sm7}
The same plots as in Fig.~\ref{sm9} for pp collisions at 7 TeV.
}
\end{figure}

In accordance with the standard picture of string decay
we use the following para\-me\-tri\-za\-tion for the pair correlation function $ \Lam{\eta,\phi}$ of a single string:
\beq \label{Lam_fit}
\Lam{\eta,\phi}=\Lambda_1 e^{-\frac{|\eta|}{\eta_1}} e^{-\frac{\fv^2}{\fv^2_1}}  +
\Lambda_2 \left(e^{-\frac{|\eta-\eta_0|}{\eta_2}}
+ e^{-\frac{|\eta+\eta_0|}{\eta_2}}\right)  e^{-\frac{(|\fv|-\pi)^2}{\fv^2_2}}  \ .
\eeq
The first term in the formula corresponds to the near-side correlation peak at $\phi=0$,
which originates from the hadronization of a given string segment.
The widths of this peak in azimuth and rapidity are characterized by the parameters $\eta_1$ and $\fv_1$,
while its amplitude is characterized by  $\Lambda_1$.
The second term in the formula corresponds to the away-side ridge-like structure at $\phi=\pi$.
This manifests as the overlap of two lower, wider symmetric humps which are parameterized
using $\Lambda_2$, $\eta_2$ and $\fv_2$.
These humps are shifted $\pm\eta_0$ in rapidity
relative to the near-side peak position.
They originate from the hadronization of two string segments on either side of the given string segment,
so the $\eta_0$ parameterizes the mean rapidity length of a string decay segment.
In the parametrization (\ref{Lam_fit}), we imply that:
\beq \label{f_obl}
|\fv|\leq\pi  \ .
\eeq
For $|\fv| > \pi$ we must periodically extend $\Lam{\eta,\phi}$ to $\fv\to\fv+2\pi k$.
With such completion the $\Lam{\eta,\phi}$ meets the following requirements:
\beq \label{Lam_sym}
\Lam{-\eta,\fv}=\Lam{\eta ,\fv} \ , \hs1
\Lam{\eta ,-\fv}=\Lam{\eta ,\fv} \ , \hs1
\Lam{\eta ,\fv+2\pi k}=\Lam{\eta ,\fv}
 \ .
\eeq

\subsection{Fitting the Model Parameters}\label{fitting}Once the single string pair correlation function $ \Lam{\eta,\phi}$ is parameterized, as in (\ref{Lam_fit}), one can calculate the FB multiplicity correlation coefficient for windows of an arbitrary width in rapidity and azimuth
using (\ref{blarg}).  In this formula the integrals $\JFB$ and $\JFF$ are given
by the equations (\ref{subst_int1}) and (\ref{subst_int2}), applicable in the mid-rapidity region.
See the technical details of the calculation, as well as the resulting formula, (\ref{bcorrR}), in the Appendix A.

To fit the model parameters we first calculated the FB multiplicity correlation strength
in the simplest case using windows
that are small both in rapidity and azimuth acceptance.
Then the experimental data for the correlation strength between two small windows situated at varying
rapidity $\yFB$ and azimuth $\fFB$ separation
was used to determine all parameters in both the SR (\ref{bSsm}) and the LR  (\ref{bLsm}) contributions.
In this we have only used the FB correlation coefficient  $b_{corr}$ defined with symmetrical FB windows.
As we have seen above, (\ref{bequal}), this implies that $b_{corr}\equiv\brel=b$.

We fit the parameters of the single string pair correlation function $\Lam{\eta ,\phi}$, (\ref{Lam_fit}),
and the scaled variance parameter $\omega_N$, (\ref{omN}), to the data presented in \cite{AltsPhD}.
This experimental data measures the FB correlation coefficient, $b_{corr}$, for charged particles
with transverse momenta 0.3$<\!\!p_T\!\!<$1.5 GeV/c,
obtained for symmetric windows of $\DyF=\DyF=\Dy=0.2$ rapidity acceptance
and $\DfF=\DfF=\Df=\pi/4$ azimuthal acceptance.

The results of the fit for pp collisions at 0.9 TeV and 7 TeV are presented in Figs.~\ref{sm9} and \ref{sm7} respectively.
The obtained values of the model parameters used at fitting are presented in the first and the last column
in Table \ref{param97}.
The data at 2.76 TeV were not analysed in \cite{AltsPhD} with windows,
that are small both in rapidity and azimuth acceptance.
So, as a rough estimate for model parameters at this energy, we have used the averaged
values of the ones at  0.9 TeV and 7 TeV.

Note that in the resulting formula, (\ref{bcorrR}), the parameters $\mu_0, \omega_N, \Lambda_1$ and $\Lambda_2$
appear only via the products $\mu_0\omega_N$, $\mu_0\Lambda_1$ and $\mu_0\Lambda_2$.
Hence, values for these products were fitted instead of the parameters themselves.
Recall, that $\mo$ is the average rapidity density of the charged particles produced by one string, (\ref{Lam2}),
and $\omega_N$ is the event-by-event scaled variance of the number of strings, (\ref{omN}).

Fig.~\ref{sm9} and Fig.~\ref{sm7} show the general behavior of the correlation coefficient  $\bcor$
for sufficiently small FB windows by using $\Lam{\eta,\phi}$, (\ref{Lam_fit}), as predicted in the string fragmentation.
These plots show a large, narrow, near-side peak at $\yFB=0$, $\fFB=0$, as well as the shorter,
wider away-side hump at $\fFB=\pi$.
These peaks are elevated by the constant term corresponding to the LR correlation,  (\ref{bLsm}),
which originates from the event-by-event fluctuation in the number of emitting sources.

In Table \ref{param97} we see that as the energy increases from 0.9 TeV to 7 TeV, the value of $\mu_0\omega_N$ triples,
while the parameters that characterize the single string pair correlation function $\Lam{\eta,\phi}$, (\ref{Lam_fit}),
do not change considerably.
With the increase of energy, the near-side peak becomes a little bit narrower and higher,
while the shape of the away-side ridge-like structure parameters remaining the same.

\begin{table}[tb]
  \centering
\begin{tabular}{|c|c|c|c|c|}
 \hline
 \multicolumn{2}{|c|}{$\sqrt{s}$,\  TeV}&0.9&2.76&7.0 \\
 \hline  \hline
 LRC &$\mu_0\omega_N$&0.7 & 1.4 &2.1\\
 \hline \hline
 &$\mu_0\Lambda_1$&1.5 & 1.9 & 2.3 \\
 &$\eta_1$&0.75 &0.75&0.75 \\
 &$\phi_1$&1.2 &1.15&1.1 \\
 \cline{2-5}
 SRC&$\mu_0\Lambda_2$&0.4 &0.4&0.4 \\
 &$\eta_2$&2.0 &2.0&2.0 \\
 &$\phi_2$&1.7 &1.7&1.7 \\
 \cline{2-5}
 &$\eta_0$&0.9 &0.9&0.9 \\
 \hline
\end{tabular}
 \caption[dummy]{\label{param97}
The parameters of the model
used at the comparison with the experimental data \cite{AltsPhD,PoS12Feof},
see formulae (\ref{omN}), (\ref{blarg}) and (\ref{Lam_fit}).
}
\end{table}

\section{Comparison with the Experimental Data}In previous section we have fixed all model parameters by the data
on the FB correlation coefficient with small acceptance windows,
see the Table \ref{param97}.
So now,
basing on the formulae (\ref{subst_int1}), (\ref{subst_int2}) and (\ref{blarg}),
we can calculate
the values of the FB correlation coefficient $b_{corr}$ for large acceptance windows,
within which $\Lam{\eta ,\fv}$ cannot be treated as constant.
Note, no additional free parameters are introduced in this process.

In this section we
calculate the values of the FB correlation coefficient $b_{corr}$ for symmetric windows
separated only in rapidity, i.e. with the full $2\pi$ azimuthal acceptance.
This case is commonly used in experiment.
We study the dependence of $b_{corr}$ on the windows' acceptance widths and
on the rapidity separation size.
Note that for this case $b_{corr}$ is given by (\ref{blarg}), in which $\JFB$ and $\JFF$ are given
by (\ref{subst_int3})--(\ref{Lam-y}).
For details see Appendix A, namely formulae (\ref{int1c})--(\ref{bcorrR}).

\begin{figure}[t]
\centerline{
\includegraphics[width=80mm,angle=0]{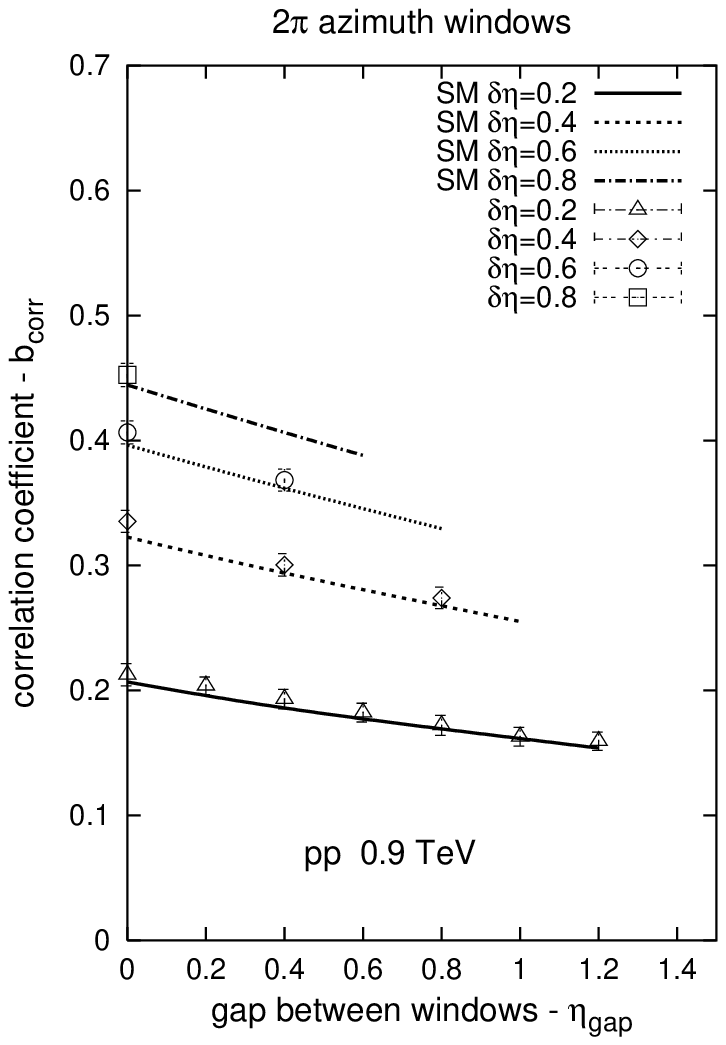}
\includegraphics[width=80mm,angle=0]{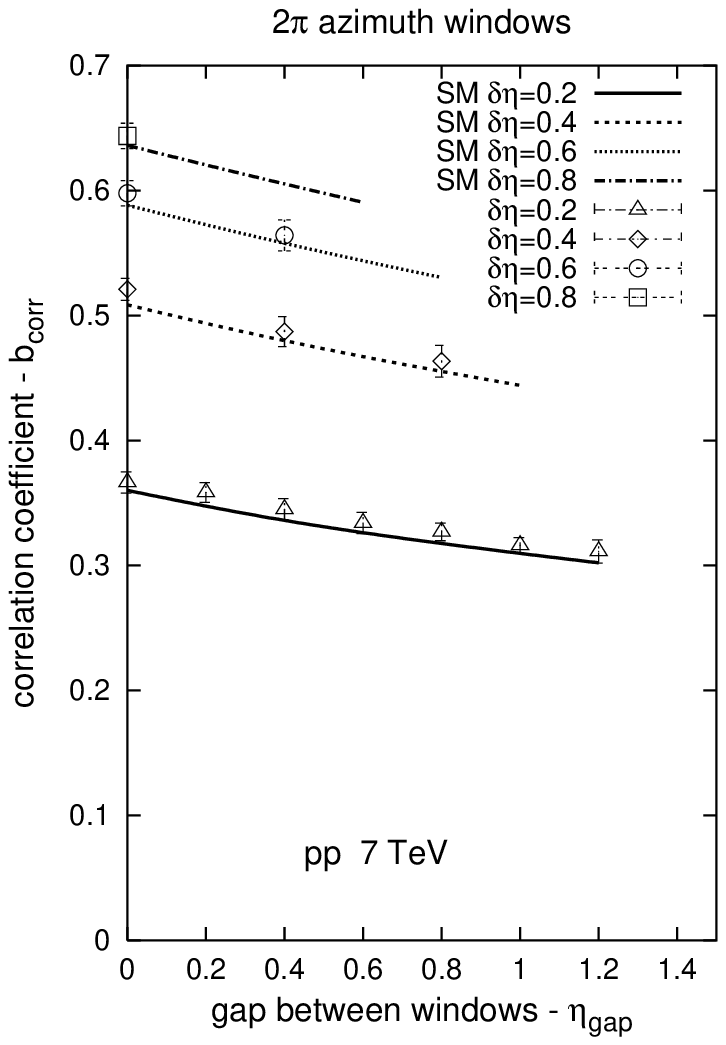}
}
\caption{\label{larg97}
The forward-backward (FB) correlation coefficient, (\ref{defb}), in pp collisions at 0.9 TeV and 7 TeV for symmetric windows with $\DyF=\DyB=\Dy=0.2$, 0.4, 0.6 and 0.8 rapidity acceptance and full 2$\pi$-azimuthal acceptance.
Its value is plotted as a function of
rapidity gap between windows,
$\eta^{}_{gap}\equiv\yFB-\Dy$,  (\ref{gapsym}).
The curves are the results of calculations by the model
that treats strings as independent identical emitters (SM), according to the formulae
(\ref{subst_int1}), (\ref{subst_int2})  and (\ref{blarg}).
The values of the parameters are shown in Table \ref{param97}.
The corresponding experimental data are taken from the preliminary results of ALICE \cite{PoS12Feof}.
}
\end{figure}

\begin{figure}[t]
\centerline{
\includegraphics[width=80mm,angle=0]{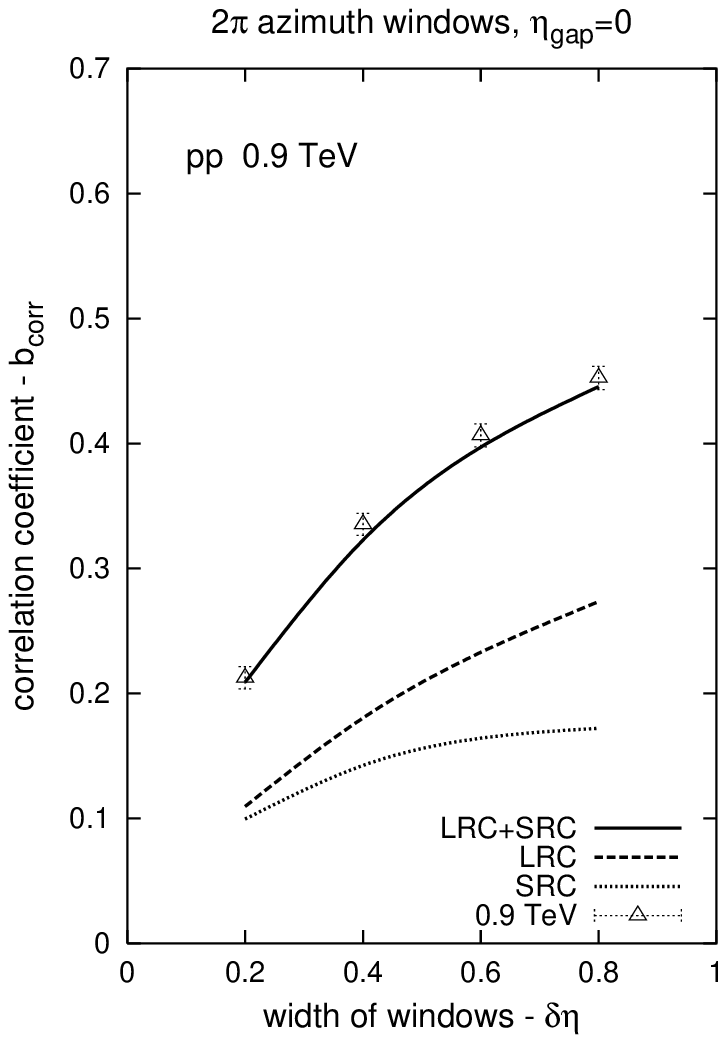}
\includegraphics[width=80mm,angle=0]{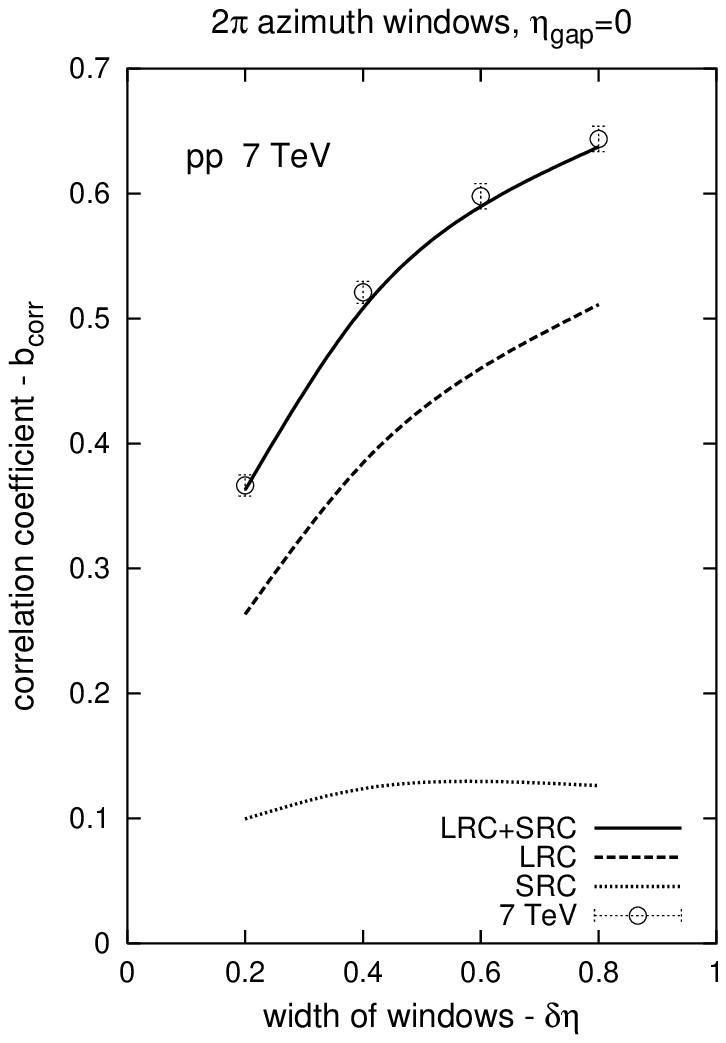}
}
\caption{\label{larg97dy}
The same results as in Fig.~\ref{larg97}, now  as a function of windows width, $\DyF=\DyB=\Dy$,
at zero gap between the windows, $\eta^{}_{gap}=0$.
Additionally, the relative contributions of the long-range (LRC), (\ref{bLsm}),
and short-range correlations (SRC), (\ref{bSsm}),
resulting from the model
that treats strings as independent identical emitters (SM), (\ref{blarg}),
are shown.}
\end{figure}

\begin{figure}[t]
\centerline{
\includegraphics[width=80mm,angle=0]{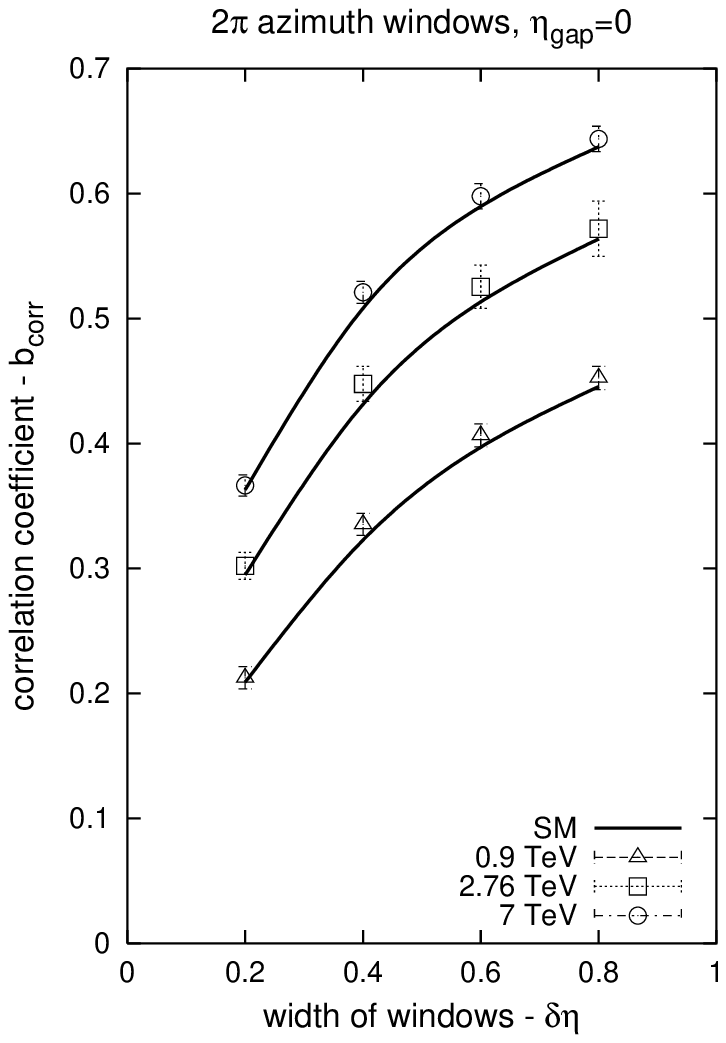}
}
\caption{\label{lar97dy}
The FB correlation coefficient results as in Fig.~\ref{larg97dy},
now the results of the 0.9, 2.76 and 7 TeV calculations and data are on the same plot for comparison.
Note both the long-range and short-range contributions are taken into account here.
As before, the corresponding experimental data are taken from the preliminary results of ALICE \cite{PoS12Feof}.
}
\end{figure}

The results for pp collisions at 0.9, 2.76 and 7 TeV with FB window widths ranging from 0.2 to 0.8 rapidity units
are presented in Figs.~\ref{larg97}-\ref{lar97dy}.
Fig.~\ref{larg97} shows the results as a function of the rapidity gap between windows, $\eta^{}_{gap}\equiv\yFB-\Dy$,
as defined in (\ref{gapsym}). Figs.~\ref{larg97dy} and \ref{lar97dy} show the results as a function of window width, $\DyF=\DyF=\Dy$,
at zero rapidity separation, $\eta^{}_{gap}=0$.
These calculated results are plotted against the preliminary experimental measurements for the $b_{corr}$
for charged particles with transverse momenta 0.3$<\!\!p_T\!\!<$1.5 GeV/c in ALICE \cite{PoS12Feof}.

Figs.~\ref{larg97}-\ref{lar97dy} show a good agreement between the values calculated by the string model (SM) with strings
treated as independent identical emitters, (\ref{blarg})-(\ref{Lam-y}), and the experimental data for FB correlations in large,
$2\pi$-azimu\-th\-al windows \cite{PoS12Feof}.
Note that the agreement was achieved without using any additional adjusting parameters;
all model parameters were fixed by the experimental data for small windows separated in azimuth and rapidity \cite{AltsPhD},
as discussed in Sec.~\ref{fitting} and shown in Table \ref{param97}.

Fig.~\ref{larg97dy} shows that the
relative contribution of the long-range correlation (LRC), (\ref{bLsm}), is considerably larger at 7 TeV
when compared to 0.9 TeV. This increase reflects the significant growth of the event-by-event fluctuations
of the number of particle emitting sources, i.e. strings, with energy.  This fluctuation is  quantified
by the scaled variance parameter $\omN$,  (\ref{omN}), in Table \ref{param97}.
At the same time, the contribution of the short-range correlation (SRC), as in (\ref{bSsm}),
which characterizes the properties of a single source, remains practically the same across both energy scales.


\section{Alternative Observables} \label{alt} In this section we would like to discuss the introduction of more suitable observables for the future FB correlation studies.

Equations (\ref{smallmF-r})--(\ref{bet_sm-r}) show that if the acceptance of small symmetric FB windows goes to zero,
$\aF=\aB\to 0$, then all traditional FB correlation coefficients $b$ , $\brel$, and $\bsym$,
as defined in formulae (\ref{defb}), (\ref{brel}), and (\ref{defbsym}), also vanish.
This unpleasant dependence of the correlation coefficient on the window width is a consequence of using
the variance $D_\nF$, in the definition of $b$, as in (\ref{defb}).
Explicitly, it was demonstrated in (\ref{nBF}) and (\ref{smallmF-r})--(\ref{D_sm-r}),
that $\av{\nF\nB}-\av{\nF}\av{\nB}\sim\aF\aB$ and $D_\nF\sim\aF$, which is a model-independent consequence of this limit.
We can eliminate this drawback if we redefine $b$ in (\ref{defb}) by normalizing the correlator,
$\av{\nF\nB}-\av{\nF}\av{\nB}$, by the product $\av{\nF}\av{\nB}$ instead of $D_\nF$.
Hence, we can  introduce the observable
\beq \label{bm-def}
\bm\equiv \frac{\av{\nF\nB}-\av{\nF}\av{\nB}}{\av{\nF}\av{\nB}}
= \avL{\nFr\nBr}-1    \ .
\eeq
For sufficiently small windows in both rapidity and azimuth, (\ref{nBF}) and (\ref{IBFsm-r}) can be used to write:
\beq \label{bmC_2}
\bm=C_2(\yF,\yB;\fFB)    \ .
\eeq
For windows with small rapidity acceptance, yet large azimuthal acceptance, $\DfF=\DfB=2\pi$, we have:
\beq \label{bmC2}
\bm= C_2(\yF,\yB)=\frac{1}{\pi}\int_{0}^{\pi} \!\!d\fv \,  C_2(\yF,\yB;\fv)   \ ,
\eeq
where we used (\ref{rofF2})--(\ref{C2}).

So, as the window acceptance goes to zero, $\aF=\aB\to 0$, the new observable $\bm$, (\ref{bm-def}), results
in a finite, nonzero two-particle correlation function.
This is in clear contrast with the properties of $b$, $\brel$ and $\bsym$.

We should note that the traditionally defined, (\ref{defb}), correlation coefficient $b$ is also proportional
to the two-particle correlation function  $C_2(\yF,\yB;\fFB)$, as in (\ref{bet_sm-r}), but the proportionality factor depends
on the width of windows and goes to zero at $\aF=\aB\to 0$.

Working from (\ref{D_sm-r}), another possible redefinition is to use the differences $D_\nF-\av\nF$ and $D_\nB-\av\nB$ instead of  $D_\nF$ and $D_\nB$ to normalize in (\ref{defbsym}).  Hence, we introduce:
\beq \label{brob}
\brob\equiv \frac{\av{\nF\nB}-\av{\nF}\av{\nB}}{\sqrt{D_\nF-\av{\nF}}\sqrt{D_\nB-\av{\nB}}}   \ .
\eeq
As before, for sufficiently small windows in both in rapidity and azimuth, (\ref{IBFsm-r}) and (\ref{D_sm-r}) can be used to write:
\beq \label{brobC_2}
\brob=\frac{C_2(\yF,\yB;\fFB)}{\sqrt{C_2(\yF,\yF;0) C_2(\yB,\yB;0)}} \ .
\eeq
If these windows are in the mid-rapidity region, (\ref{brobC_2}) reduces to
\beq \label{brobC_2t}
\brob=\frac{C_2(\yFB,\fFB)}{C_2(0,0)} \ .
\eeq
For windows with small rapidity acceptance, yet large azimuthal acceptance, $\DfF=\DfB=2\pi$, we have:
\beq \label{brobC2}
\brob= \frac{C_2(\yF,\yB)}{\sqrt{C_2(\yF,\yF) C_2(\yB,\yB)}}   \ ,
\eeq
where $C_2(\yF,\yB)$ is defined by (\ref{C2}).  Again, if these windows are in the mid-rapidity region,
(\ref{brobC2}) can be simplified to
\beq \label{brobC2t}
\brob=\frac{C_2(\yFB)}{C_2(0)}
\eeq
We see that $\brob$ has a finite limit at small window acceptances, similarly to the defined above $\bm$.

Note that the definition of $\brob$, (\ref{brob}), is closely connected with ``robust variance" \cite{Voloshin02}, as defined by:
\beq \label{robust}
R_n=
\frac{D_n-\av{n}}{\av{n}^2_{}}    \ .
\eeq
Using (\ref{bm-def}) and (\ref{brob}),  we can write:
\beq \label{brobR}
\brob= \frac{\bm}{\sqrt{R_\nF R_\nB}}      \ .
\eeq

According to the model described in Sec.~\ref{model},
when describing the FB correlations with small observation windows,
these alternative observables $\bm$ and $\brob$,  (\ref{bm-def}) and (\ref{brob}), become:
\beq \label{bmod-sm}
\bm
=\frac{\omega_N+ \Lam{\yF,\yB;\fFB}}{\av N}
\ ,
\eeq
\beq \label{brob-sm}
\brob=\frac{\omega_N+ \Lam{\yF,\yB;\fFB}}{\omega_N+ \Lam{\yF,\yF;0}}
\ .
\eeq
Again, it is clear that these new observables have finite, non-zero limits at small acceptances $\aF$, $\aB$.
Additionally, they are described by the single string pair correlation function, $\Lam{\yF,\yB;\fFB}$,
the mean number of string, $\av{N}$, and the scaled variance of this number, $\omega_N=D_N/\av N$,
as defined in (\ref{omN}).

Note (\ref{bmod-sm}) states that $\bm$ is inversely proportional $\av{N}$,
such that  if the mean number of sources is large, e.g. as in AA collisions,
it may be preferable to use as a correlation measure instead of $\bm$ the product:
\beq \label{bmp}
\rho_1^{}(0)\,\bm  =\left.\frac{dN_{ch}}{d\eta}\right|_{\eta=0}\!\!\!\!\cdot C_2(\yF,\yB;\fFB)
= \mo\omega_N+ \mo\Lam{\yF,\yB;\fFB} \ .
\eeq
Note that we have used $\rho_1(0) = \mo\,\av N$, where the $\mo$ is the density in rapidity of charged particles produced by one string, (\ref{Lam2}).

In comparing the different definitions of the multiplicity correlation coefficient, it is clear
that the traditional definitions (\ref{defb}), (\ref{brel}) and (\ref{defbsym}) of the FB correlation coefficient
lead to a strong dependence on the window acceptance size.
This causes the correlation coefficient to go to zero with the window acceptance.
This also means that results obtained for windows of different widths cannot be compared directly.
Hence, it may be preferable to use the newly proposed observables (\ref{bm-def}) and (\ref{brob})
in future FB correlation studies, as they remain nonzero in the limit that the window acceptance goes to zero.

\section{Conclusions} We have extended the traditional FB multiplicity correlation analysis
by allowing for acceptance windows separated both in rapidity and azimuth.
We have shown that this can help distinguish the two main contributions to the correlation strength.
The first of these contributions arises from the event-by-event fluctuation in the number of sources,
while the second originates to the single source pair correlation function.

In the mid-rapidity region, the first contribution
does not depend on the separation between the windows in rapidity and azimuth, which leads
to long range (LR) correlations between window multiplicities.
This LR contribution was demonstrated to be proportional
to the scaled event-by-event variance of the number of sources, $\omN$.
The analysis of this contribution could hence provide an important, quantitative, physical channel to  measure
the magnitude of this fluctuation in a given process.

The second contribution is distinct, as it originates from the correlation between multiplicities produced by a single source.
It can arise from different physical processes such as
the formation and decay of clusters,  resonances or minijets during the string fragmentation.
Its value, (\ref{bSsm}), depends on the separation between the backward and forward windows
in both rapidity $\yFB$ and azimuth $\fFB$.
This contribution is proportional to the single source pair correlation function,  $\Lambda(\yFB,\fFB)$.
It decreases to zero at large separation between windows,
which means it only contributes to the short range (SR) correlations.
This contribution's dependence on $\yFB$ and $\fFB$ can be extracted from experimental measurements
of the multiplicity correlation between sufficiently small FB windows in rapidity and azimuth,
as in Figs.~\ref{sm9} and \ref{sm7}, and Table \ref{param97}.
We also noted, as in (\ref{D_mF}),
that the presence of this SR correlation necessitates that strings are non-poissonian emitters.

By comparing the different definitions of the multiplicity correlation coefficient we noted that
the traditional definitions (\ref{defb}), (\ref{brel}) and (\ref{defbsym}) of the FB correlation coefficient
led to a strong dependence on the acceptance of the windows, with the correlation coefficient going
to zero with the acceptance.
Hence,  the results obtained for the windows of different width cannot be compared directly.
As a solution, we proposed suitable observables in (\ref{bm-def}) and (\ref{brob}) for the future FB correlation studies,
which have a nonzero limit as the acceptance goes to zero.
The strong non-linear dependence of the traditionally defined FB correlation coefficient
on the width of the windows and on the value of gap between them, is well described in the framework
of the model with strings as independent identical sources (see Figs. \ref{larg97}--\ref{lar97dy}).

Using a model independent method, we have shown that the two-particle correlation function, $C_2$,
can be determined by measuring the FB multiplicity correlation coefficient  between two small windows
separated in rapidity and azimuth.  This still holds when the particle distribution in rapidity is not flat,
e.g. in pA interactions, and $C_2(\eta_1, \eta_2; \Delta\phi)$ does not only depend on the difference
of rapidities, $\Delta\eta=\eta_1-\eta_2$, but both on $\eta_1$ and $\eta_2$.

It is worth noting that this approach does not need to use an event mixing procedure, as applied in the
di-hadron correlation analysis,
in which one assumes from the very beginning the dependence of two-particle correlation function
only on the differences $\Delta\eta$ and $\Delta\phi$.
Even in a mid-rapidity region, where the application of the di-hadron correlation approach is justified,
the results obtained by this method depend on the details of track and/or event mixing used in the approach
for the imitation of the uncorrelated particle production, see Appendix C for details.
This leads to the uncertainty in determination
of the common pedestal in $C_2$ by the di-hadron correlation analysis
and hence, as it was demonstrated, to the loss
of important physical information
on fluctuations in the number of sources.

\section*{Acknowledgements}
The author thanks M.A.~Braun, G.A.~Feofilov, and I.~Altsybeev for useful discussions
as well as E.~Gillies for helping to prepare the manuscript.
The work was supported by the RFBR grant 12-02-00356-a and
the Saint-Petersburg State University grant 11.38.197.2014.



\appendix

\section{Calculation of the Integrals Over Rapidity and Azimuth Windows}
\label{apA}
\begin{figure}[t]
\centerline{
\includegraphics[width=80mm,angle=0]{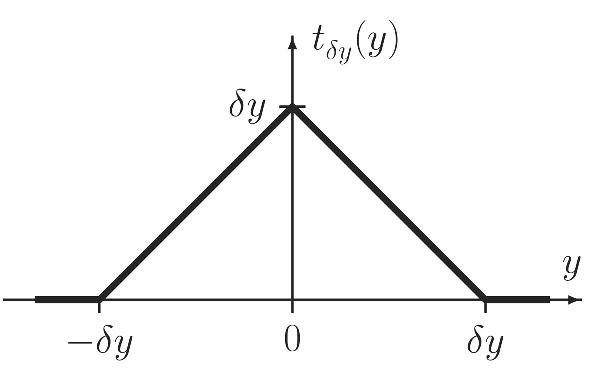}
}
\caption[dummy]{\label{triang}
The phase space ``triangular" weight function,
arising at integration over non-periodic FB windows, as in (\ref{tri}).}
\end{figure}

For symmetric rapidity windows, $\DyB=\DyF=\Dy$, whose centers are separated by $\yFB=\eta_F-\eta_B$, one has:
\beq \label{Int_y}
\int_\DyF \!\!d\eta_1 \int_\DyB \!\!d\eta_2\,  f(|\eta_1-\eta_2|)= \int_{-\Dy}^\Dy \!\!d\eta\,  f(|\yFB+\eta|) \,t_\Dy(\eta)  \ ,
\eeq
where $t_{\delta y}(y)$ is the ``triangular" weight function, see Fig.~\ref{triang}:
\beq \label{tri}
t_{\delta y}(y)= [\theta(-y)(\delta y+y) +\theta(y)(\delta y-y)]\,\theta(\delta y-|y|) \ .
\eeq
Formula (\ref{Int_y}) is valid for any distance between the centers of windows,
in particular for coinciding windows, for which $\yFB=0$.  In this case, we have:
\beq \label{Int_yF}
\int_\DyF \!\!d\eta_1 \int_\DyF \!\!d\eta_2\,  f(|\eta_1-\eta_2|)= \int_{-\Dy}^\Dy \!\!d\eta\,  f(|\eta|) \,t_\Dy(\eta)
= 2\int_{0}^\Dy \!\!d\eta\,  f(|\eta|) (\Dy-\eta)  \ .
\eeq
The same general formula can be used for the integration over azimuthal windows:
\beq \label{Int_fF}
\int_\DfF \!\!d\fv_1 \int_\DfB \!\!d\fv_2\,  f(|\fv_1-\fv_2|)= \int_{-\Df}^\Df \!\! d\fv \,   f(|\fFB+\phi|) \,t_{\Df}(\phi) \ .
\eeq
In this case the function $f(|\phi|)$ is periodic in $\phi$ such that $f(|\phi|)=f(|\phi+2\pi k|)$.
This implies that windows of full azimuthal acceptance, i.e. the full $2\pi$ range, allow for (\ref{Int_fF})
to be simplified to the following:
\beq \label{Int_2pi}
\int_{-2\pi}^{2\pi} \!\! d\fv \,  f(|\fFB+\phi|) \,t_{2\pi}(\phi)=
4\pi\int_{0}^{\pi} \!\! d\fv \,  f(|\phi|)
  \ .
\eeq

Recalling the general form of $\JFF$ and $\JFB$ in formulae (\ref{subst_int1}) and (\ref{subst_int2}),
(\ref{Int_y})--(\ref{Int_yF}) imply that these can be simplified for large symmetric windows
in the central rapidity region to:
\beq \label{sub_int1}
\JFB=(\Dy\Df)^{-2} \int_{-\Dy}^\Dy \!\!d\eta \int_{-\Df}^\Df \!\! d\fv \,  \Lam{\yFB+y,\fFB+\phi}
\,t_\Dy(\eta)\,t_\Df(\phi)   \ ,
\eeq
\beq \label{sub_int2}
\JFF=4(\Dy\Df)^{-2} \int_{0}^\Dy \!\!d\eta \int_{0}^\Df \!\! d\fv \,  \Lam{\eta ,\fv}
\, (\Dy-y)  \, (\Df-\phi)  \ ,
\eeq
 where $\Dy$ and $\Df$ are the widths of the observation windows, while $\yFB$ and $\fFB$
are the corresponding distances between their centers.
Note that this result implies that $\Lam{\eta ,\fv}$ satisfies the conditions in (\ref{Lam_sym}).
A similar procedure can be applied to integrals $\IBF$ (\ref{IBFsm})  and $\IFF$ (\ref{IFFsm})
in Sec.~\ref{twopart}, yielding an analogous result.

To further simplify the numerical calculations, it is important to note that the dependencies on $\eta$ and $\phi$
in near-side and away-side contributions factorize.  This arises in the pair correlation function
$ \Lam{\eta,\phi}$, given by the formulae (\ref{Lam_fit}), which is of the form:
\beq \label{Lam_fact}
\Lam{\eta,\phi}=\sum_{i=1}^2 \Lambda_i \, F_i(\eta) f_i(\fv)  \ ,
\eeq
such that $i=1$ corresponds to the near-side contribution, while $i=2$ to the away-side contribution:
\beq \label{F1}
F_1(\eta)=e^{-\frac{|\eta|}{\eta_1}}  \ ,    \hs1
F_2(\eta)=e^{-\frac{|\eta-\eta_0|}{\eta_2}} + e^{-\frac{|\eta+\eta_0|}{\eta_2}}   \ ,
\eeq
\beq \label{f1}
f_1(\fv)= e^{-\frac{\fv^2}{\fv^2_1}} \ , \hs1
f_2(\fv)=
e^{-\frac{(|\fv|-\pi)^2}{\fv^2_2}}     \ .
\eeq
In this case
the integrals  (\ref{sub_int1}), (\ref{sub_int2}), and hence the resulting FB correlation coefficient $b_{corr}\equiv\brel=b$, (\ref{blarg}), can be expressed through the  one dimensional integrals:
\beq \label{int1c}
J_{FB}(\yFB ,\fFB )=\sum_{i=1}^2 \Lambda_i \, H_i(\yFB ) h_i(\fFB )  \ , \hs1
J_{FF}=\sum_{i=1}^2 \Lambda_i \, H_i(0) h_i(0)  \ ,
\eeq
where (\ref{sub_int1}) implies:
\beq \label{H}
H_i(\yFB )=  (\Dy)^{-2}
\int_{-\Dy}^\Dy \!\!d\eta \,  F_i(\eta+\yFB) \,t_\Dy(\eta) \ ,
\eeq
\beq \label{h}
h_i(\fFB )=  (\Df)^{-2}
\int_{-\Df}^\Df \!\! d\fv \, f_i(\phi+\fFB) \,t_\Df(\phi)    \ ,
\eeq
and (\ref{sub_int2}) implies:
\beq \label{H0}
H_i(0)=  2(\Dy)^{-2}
\int_{0}^\Dy \!\!d\eta \,  F_i(\eta) \, (\Dy-\eta) \ ,
\eeq
\beq \label{h0}
h_i(0)=  2(\Df)^{-2}
\int_{0}^\Df \!\! d\fv \, f_i(\phi) \, (\Df-\phi)     \ .
\eeq

Note that we must ensure parity and azimuthal periodicity of $f_1(\fv)$ and $f_2(\fv)$,
as in (\ref{f_obl}) and (\ref{Lam_sym}).
Then for $2\pi$-azimuth windows,
using these relations, along with (\ref{Int_2pi}), the formulae (\ref{h}) and (\ref{h0}) can be further simplified to:
\beq \label{h2pi}
h_i(\fFB )= h_i(0)= \frac{1}{\pi}
\int_{0}^{\pi} \!\! d\fv \, f_i(\phi)    \ .
\eeq

Substituting  (\ref{int1c})
into (\ref{blarg}), we obtain an expression for the FB correlation coefficient in terms of the factorized string pair correlation function, $ \Lam{\eta,\phi}$ in (\ref{Lam_fact}).  This yields:
\beq\label{bcorrR}
b_{corr} =
\frac{[\omega_N+ \sum_{i=1}^2 \Lambda_i \, H_i(\yFB ) h_i(\fFB )]\mo \aF}
{1+[\omega_N+ \sum_{i=1}^2 \Lambda_i \, H_i(0) h_i(0)]\mo \aF}    \ .
\eeq

\section{Connection of the Correlator and Variance with the Ones of a Single Source}
In the two stage model in \cite{PLB00,EPJC04,Vestn1}, the assumption that each string contributes
to the particle production in both FB windows allows the observable correlator and variance,
$\av{\nF\nB}-\av{\nF}\av{\nB}$ and $D_\nF$, to be expressed through the correlator and variance of one source,
 $\omBF-\omF\omB$ and $D_\mF$.
This is given by:
\beq \label{corcor}
\av{\nF\nB}-\av{\nF}\av{\nB}
= \av N (\omBF-\omF\omB)+D_N \omF\omB
   \ ,
\eeq
\beq \label{DD}
D_\nF=\av N D_\mF +D_N \omFF
   \ ,
\eeq
where $D_N=\av {N^2}-\av N^2$ and $\av N$ are the event-by-event variance and the mean number of sources,
respectively.
See \cite{Dub10} for a derivation.

For sufficiently small FB windows in both rapidity and azimuth we have seen in (\ref{C2ex}) that
\beq \label{C2-cor}
C_2(\yF,\yB;\fFB)=\frac{\av{\nF\nB}-\av{\nF}\av{\nB}}{\av{\nF}\av{\nB}}  \ .
\eeq
This relation can also be obtained using rearranging (\ref{nBF}) and substituting in (\ref{IBFsm-r}).
Analogously,  (\ref{mBF}) and (\ref{JFBsm1}) imply for the single string pair correlation function $\Lam{\yB,\yF;\fFB}$:
\beq \label{Lam-cor}
\Lam{\yB,\yF;\fFB}
= \frac{\av{\mF\mB}-\av{\mF}\av{\mB}}{\av{\mF}\av{\mB}} \ .
\eeq
Then combining the formulae (\ref{corcor})--(\ref{Lam-cor}) and taking into account that
\beq \label{romo}
\av{\nF}=\av{N} \av{\mF} \ , \ \  \av{\nB}=\av{N} \av{\mB}  \ ,
\eeq
we again recover the formula (\ref{C2_Lam}) of the text:
$$C_2(\eta_F,\eta_B;\fv)=(\omega_N+ \Lam{\eta _F,\eta_B;\fv})/\av N \ .$$

Note that for LR correlations, the FB observation windows are separated by a large  rapidity gap, which allows one to neglect the correlations produced from the same source.  For this case, (\ref{corcor}) implies:
\beq \label{corcorLR}
\av{\nF\nB}-\av{\nF}\av{\nB} = D_N \omF\omB
 \ .
\eeq
Hence (\ref{DD}) and (\ref{corcorLR}) lead to the following expression for the LR correlation coefficient, as defined in (\ref{brel}):
\beq \label{brelLR}
\brelLR =\frac{a} {1+a} \ ,  \hs 1  a=\frac{\omN}{\omega_{\mF}}\omF     \ ,
\eeq
where $\omN$ and $\omega_{\mF}$  are the corresponding normalized variances:
\beq \label{ommuF}
\omN=\frac{D_N}{\av{N}} \ ,  \hs 1 \omega_{\mF}=\frac{D_\mF}{\omF} \ .
\eeq
The expression given in (\ref{brelLR}) coincides with the expression for the LR
correlation coefficient obtained in \cite{Dub10}.

Substituting  the expression (\ref{D_mF}) for $D_\mF$
into (\ref{brelLR}), one obtains:
\beq \label{blargLR}
\brelLR =\frac{\av\mF\omega_N}{1+\av\mF[\omega_N+ \JFF]}    \ ,
\eeq
which coincides with the LR contribution to the FB correlation coefficient in (\ref{blarg}). Recall that for sufficiently small FB windows in rapidity and azimuth, (\ref{JFFsm}) implies $\JFF=  \Lam{\yF,\yF;0}$.

\section{Connection with the Untriggered  Di-Had\-r\-on Correlation Approach}
In di-hadron correlation analysis, the following alternative definition of the two-particle correlation
function $C$ is used
\cite{STARdihadr09,CMSridge10}:
\beq \label{C-SB}
C=\frac{S}{B}-1    \ ,
\eeq
where
\beq \label{S}
S=\frac{d^2 N}{d\Delta \eta\  d\Delta\phi}    \ .
\eeq
Here $\Delta \eta=\eta_1-\eta_2$ and $\Delta\phi=\fv_1-\fv_2$ are the distances between two particles
in rapidity and in azimuth.
One then takes into account all possible pair combinations of particles produced
in a given event over a large rapidity interval $\Delta\eta\in(Y_1,Y_2)$.
$B$ is the same as $S$, but obtained using the event mixing procedure
to imitate
uncorrelated particle production.

In contrast with (\ref{C_2_1}), this definition implies from the very beginning that the translation invariance in rapidity holds and hence that the result depends
only on $\Delta \eta=\eta_1-\eta_2$ for any $\eta_1,\eta_2 \in (Y_1,Y_2)$.
Specifically, all the pairs with the same difference $\eta_1-\eta_2$ contribute to the same bin of the multiplicity distribution, irrespective of
their average, $(\eta_1+\eta_2)/2$. See also the discussion in \cite{Bzdak12} for more details.

This assumption is reasonable only in the central rapidity region at high energies.
Namely, it implies that in the interval $(Y_1,Y_2)$:
\beq \label{ro-trans}
\ro{\eta }=\rhoo    \ , \hs1
 \rho^{}_2(\eta_1,\eta_2;\fv)= \rho^{}_2(\eta_1-\eta_2,\fv)   \ ,
\eeq
as in formula (\ref{C_2_2}).
In this case we have  for the numerator in (\ref{C-SB}):
\beq \label{S1}
S(\Delta \eta,\Delta\fv)=  \int_{Y_1}^{Y_2} d\eta _1 d\eta _2\, \rho^{}_2(\eta_1-\eta_2,\Delta\fv)\, \delta(\eta_1-\eta_2-\Delta \eta)
\eeq
or for the commonly used symmetric interval $(-Y/2,Y/2)$:
\beq \label{S2}
S(\Delta \eta,\Delta\fv)= \rho^{}_2(\Delta \eta,\Delta\fv)\,  t_Y\!(\Delta \eta)   \ ,
\eeq
where  the  $t_Y(\Delta \eta)$ is  the ``triangular" weight function (\ref{tri}),
defined in Appendix A and shown in Fig.~\ref{triang}.

For mixed events, $\rho^{}_2(\eta_1,\eta_2;\Delta\fv)$ in the denominator of (\ref{C-SB}) should  be replaced the by the product $\rho^{}_1(\eta_1) \rho^{}_1(\eta_2)$. Due to the translation invariance in rapidity, this product simplifies to $\rho^{2}_0$.  Hence:
\beq \label{B2}
B(\Delta \eta,\Delta\fv) = \rho^{2}_0 \   t_Y\!(\Delta \eta) \ .
\eeq
Substituting into (\ref{C-SB}) we get
\beq \label{C-C2}
C(\Delta \eta,\Delta\fv)=   \frac{\rho^{}_2(\Delta \eta,\Delta\fv)}{\rho^{2}_0}-1=C^{}_2(\Delta \eta,\Delta\fv) \ ,
\eeq
where we have taken into account (\ref{C_2_1}) and (\ref{C_2_2}).
We see that if translational invariance in rapidity holds within the interval $(Y_1,Y_2)$,
then the definition (\ref{C-SB}) is equivalent to the standard one  (\ref{C_2_1}).

The drawback of this approach is that it assumes translational invariance in rapidity from the beginning.
This means it cannot be used to experimentally determine the two-particle correlation function $C_2$
for asymmetrical processes, such as pA-interactions, nor at large rapidity distances,
where translational invariance is not valid.
Moreover, (\ref{bet_sm-r}), (\ref{bmC_2}) and (\ref{brobC_2}) show that approaches based
on the analysis of the standard (\ref{defb}) or modified (\ref{bm-def}), (\ref{brob}) FB correlation
coefficients between two sufficiently small windows separated in rapidity and azimuth are more robust.
They allow for the correlation function  $C_2(\eta_1,\eta_2;\phi_1-\phi_2)$ to be measured
without using of an event mixing procedure and
without  restricting to translation invariant case only.

Furthermore, using of an event mixing procedure can lead to an uncertainty
in the experimental determination of the constant term in the two-particle correlation
function $C(\Delta \eta,\Delta\fv)$,
even in central region where the definitions of the correlation functions
$C(\Delta \eta,\Delta\fv)$ (\ref{C-SB}) and $C_2(\Delta \eta,\Delta\fv)$  (\ref{C_2_1}) are equivalent
to each other (\ref{C-C2}).
The analysis explored in this paper does not suffer from this source of error.
Recalling that this constant term in two-particle correlation function
$C_2$ corresponds to the contribution from the LR correlations, as discussed in Sec.~\ref{correl},
so it is clearly important to minimize its uncertainty.

One can illustrate the origin of the uncertainty in the constant term of $C(\Delta \eta,\Delta\fv)$
using the model with strings as independent identical emitters,
described in the Sec.~\ref{model}.
Using (\ref{S2}) and (\ref{ro2-av}), the numerator and the denominator in (\ref{C-SB}) become:
\beq \label{ro2avxx}
S(\Delta \eta,\Delta\phi)
 =\av{\rho^{N}_2(\Delta \eta,\Delta\phi)}\,  t_Y^{} (\Delta \eta)
=[\av N \Lam{\Delta \eta,\Delta\phi}+ \av{N_{}^2}]\mu_0^{2}\,  t_Y^{} (\Delta \eta)   \ ,
\eeq
\beq \label{B2a}
B(\Delta \eta,\Delta\phi)
=\int_{-Y/2}^{Y/2} \!\!d\eta_1\, d\eta_2\, \av{\rho^{N}_1(\eta_1) }\av{ \rho^{N}_1(\eta_2)}\,
\delta(\eta_1-\eta_2-\Delta \eta)
= \av N^2 \mu^{2}_0 \  t_Y^{} (\Delta \eta)  .
\eeq
Then by (\ref{C-SB}) we again recover that $C(\Delta \eta,\Delta\phi)$ is equal to
$C_2(\Delta \eta,\Delta\phi)$, which is given by the equation (\ref{C2_Lam}).

But if one applies for the imitation of uncorrelated particle production an another event mixing procedure, admitting, for example, the mixing only between events with the same multiplicity (which
corresponds approximately  to the same $N$), then instead of  (\ref{B2a}) we will have
\beq \label{B2b}
B(\Delta \eta,\Delta\fv) =\int_{-Y/2}^{Y/2} d\eta_1\, d\eta_2\, \av{\rho^{N}_1(\eta_1)  \rho^{N}_1(\eta_2)}\, \delta(\eta_1-\eta_2-\Delta \eta)
= \av {N^2} \mu^{2}_0 \  t_Y^{} (\Delta \eta) \ ,
\eeq
which by (\ref{C-SB}) and (\ref{ro2avxx}) leads to
\beq \label{C-C2b}
C(\Delta \eta,\Delta\fv)=\frac{\av N}{\av {N^2}}\Lam{\Delta \eta,\Delta\fv}
\eeq
which does not correspond
to the expression (\ref{C2_Lam}), based on the standard definition (\ref{C2})
of the two-particle correlation function $C_2$.

Comparing (\ref{C-C2b}) and (\ref{C2_Lam}), we see that the resulting $C(\Delta \eta,\Delta\fv)$ in (\ref{C-C2b}) does not have the additional constant term  $\omega_N/\av N$, the common pedestal in $C_2$, reflecting the contribution of the event-by-event fluctuation of the number of sources.
Instead, (\ref{C-C2b}) is proportional to the pair correlation function of a single string $\Lam{\Delta \eta,\Delta\fv}$
and, hence it is equal to zero in the absence of the pair correlation from one string.

So the two-particle correlation function  $C(\Delta \eta,\Delta\fv)$,
obtained using the di-hadron correlation approach (\ref{C-SB}) depends on the details of
the event mixing procedure via $B$.  This mixing procedure is implemented to imitate uncorrelated particle
production.
Due to the uncertainties in the normalization factor $B$, one cannot accurately measure
a value of the constant term, i.e. the long-range component, of $C_2$.

The same effect also takes place if one uses some arbitrary, unjustified normalization procedures
for the experimental determination of $S$ and/or $B$ in formula (\ref{C-SB}),
e.g. the normalization of $S$ by the number of pairs produced in the given event, or
the normalization of $B(\Delta \eta,\Delta\fv)$ by $B(0,0)$.
As follows from the formulae (\ref{roFB-ex})--(\ref{C2ex}), determining experimentally $C_2$
one should not introduce such additional normalization factors
and has to
take into account the contributions from different events
%average all events
at given $\Delta \eta$, $\Delta\fv$ with the same weight.

Note that this long-range component of  $C_2$ can be measured unambiguously in our approach (\ref{C2ex}),
based on the studies of the FB correlations between multiplicities in windows separated in azimuth and rapidity,
which does not use any event mixture procedure, and hence avoids the associated uncertainties.


\begin{thebibliography}{99}
\bibitem{Uhlig78}
S. Uhlig, I. Derado, R. Meinke, and H. Preissner,
\emph{Observation of charged particle correlations
between the forward and backward hemispheres in pp collisions at ISR energies},
\emph{Nucl. Phys. B} {\bf 132} (1978) 15
\bibitem{UA5_83}
K. Alpgard et al.  (UA5 Collaboration),
\emph{Forward-backward multiplicity correlations in p anti-p collisions at 540 GeV}, \emph{
Phys. Lett. B} {\bf 123} (1983) 361
\bibitem{UA5_88}
R.E. Ansorge et al. (UA5 Collaboration),
\emph{Charged particle correlations in $\bar{p}p$ collisions at c.m. energies of 200 GeV,
546 GeV and 900 GeV}, \emph{Z. Phys. C} {\bf 37} (1988) 191
\bibitem{E735}
T. Alexopoulos et al. (E735 Collaboration),
\emph{Charged particle multiplicity correlations in $p\bar{p}$ collisions at $\sqrt{s}$ = 0.3 TeV to
1.8 TeV}, \emph{Phys. Lett. B} {\bf 353} (1995) 155
\bibitem{STAR09}
B.I. Abelev et al. (STAR Collaboration),
\emph{Growth of long range forward-backward multiplicity correlations
with centrality in AuAu collisions at $\sqrt{s}$ = 200 GeV}, \emph{Phys. Rev. Lett.} {\bf 103} (2009) 172301
[arXiv:0905.0237, nucl-ex]
\bibitem{ATLAS12}
G. Aad et al. (ATLAS Collaboration),
\emph{Forward-backward correlations and charged-particle
azimuthal distributions in pp interactions using the ATLAS detector}, \emph{
JHEP} 1207 (2012) 019
[arXiv:1203.3100, hep-ex]
%%%%%%%%%%%%%%%%%%%%
\bibitem{CapKr78}
A. Capella and A. Krzywicki, \emph{Unitarity corrections to short range order: long
range rapidity correlations}, \emph{Phys. Rev. D} {\bf 18} (1978) 4120
\bibitem{PRL94}
N.S. Amelin et al., \emph{Long and short range
correlations and the search of the quark gluon plasma}, \emph{Phys. Rev. Lett.} \textbf{73} (1994) 2813
\bibitem{PLB00}
M.A.~Braun, C.~Pajares and V.V.~Vechernin, \emph{On the forward-backward correlations
in a two stage scenario}, \emph{Phys. Lett. B} \textbf{493} (2000) 54
[arXiv:hep-ph/0007241]
\bibitem{EPJC04}
M.A.~Braun, R.S.~Kolevatov, C.~Pajares and V.V.~Vechernin, \emph{Correlations between multiplicities and
average transverse momentum in the percolating color strings approach}, \emph{Eur. Phys. J. C} \textbf{32} (2004) 535
[arXiv:hep-ph/0307056]
\bibitem{Brogueira07}
P. Brogueira, J. Dias de Deus and J.G. Milhano, \emph{Forward-backward rapidity
correlations at all rapidities}, \emph{Phys. Rev. C} {\bfseries 76} (2007) 064901
[arXiv:0709.3913, hep-ph]
\bibitem{ArmMcLerPajar07}
N. Armesto, L. McLerran and C. Pajares, \emph{Long range forward-backward correlations
and the color glass condensate}, \emph{Nucl. Phys. A} \textbf{781} (2007) 201
[arXiv:hep-ph/0607345]
\bibitem{YF1}
V.V.~Vechernin and R.S.~Kolevatov,  \emph{On multiplicity and transverse-momentum correlations in
collisions of ultrarelativistic ions}, \emph{Phys. Atom. Nucl.} {\bfseries 70} (2007) 1797
\bibitem{Konchak09}
V.P. Konchakovski et al., \emph{Forward-backward correlations in nucleus-nucleus collisions:
Baseline contributions from geometrical fluctuations}, \emph{Phys. Rev. C} \textbf{79} (2009) 034910
[arXiv:0812.3967, nucl-th]
\bibitem{LMcL10}
T. Lappi and L. McLerran, \emph{Long range rapidity correlations as seen
in the STAR experiment}, \emph{Nucl. Phys. A} \textbf{832} (2010) 330
[arXiv:0909.0428, hep-ph]
\bibitem{Bzdak12}
A. Bzdak, \emph{Symmetric correlations as seen in central AuAu collisions at $\sqrt{s}$=200A GeV},
\emph{Phys. Rev. C} \textbf{85} (2012) 051901
[arXiv:1108.0882, hep-ph]
\bibitem{OlszBron13}
A. Olszewski, W. Broniowski, \emph{Forward-backward multiplicity correlations in relativistic heavy-ion
collisions in a superposition approach}, \emph{Phys. Rev. C}  \textbf{88} (2013) 044913
[arXiv:1303.5280, nucl-th]
\bibitem{Vestn1}
V.V.~Vechernin, R.S.~Kolevatov, \emph{Simple cellular model of long-range multiplicity
and $p_t$ correlations in high-energy nuclear collisions}, \emph{Vestnik SPbU ser.4}, no.2 (2004) 12
[arXiv:hep-ph/0304295]
\bibitem{Dub10}
V.V. Vechernin,  \emph{Long-range rapidity correlations in the model with independent emitters},
\emph{in Proceedings of the Baldin ISHEPP XX vol.2}, JINR, Dubna (2011) 10
[arXiv:1012.0214, hep-ph]
\bibitem{PPR}
ALICE collaboration et al., \emph{ALICE: Physics Performance Report Volume II},
\emph{J. Phys. G} {\bfseries 32} (2006) 1295
[Section: 6.5.15 - Long-range correlations, p.1749]
\bibitem{Voloshin02}
C. Pruneau, S. Gavin, and S. Voloshin, \emph{Methods for the study of particle production fluctuations},
\emph{Phys. Rev. C} {\bfseries 66} (2002) 044904
[arXiv:nucl-ex/0204011]
\bibitem{STARdihadr09}
B.I. Abelev et al. (STAR Collaboration), \emph{Long range rapidity correlations and jet production
in high energy nuclear collisions}, \emph{Phys. Rev. C} {\bf 80} (2009) 064912
[arXiv:0909.0191, nucl-ex]
\bibitem{CMSridge10}
CMS Collaboration, \emph{Observation of long-range near-side angular correlations
in proton-proton collisions at the LHC},
\emph{JHEP} 1009 (2010) 091
[arXiv:1009.4122, hep-ex]
\bibitem{PRC11}
V.V. Vechernin and H.S. Nguyen,  \emph{Fluctuations of the number of participants and binary collisions
in AA interactions at fixed centrality in the Glauber approach},
 \emph{Phys. Rev. C} \textbf{84}  (2011) 054909
[arXiv:1102.2582, hep-ph]
\bibitem{AltsPhD}
I.G.~Altsybeev, \emph{Rapidity and azimuth topology of the correlations between charge particle yields
in pp and Pb-Pb collisions in ALICE experiment at LHC},
PhD Thesis, Saint-Petersburg State University, Saint-Petersburg, 2013.
\bibitem{PoS12Feof}
G.A.~Feofilov, et al. (for ALICE collaboration),  \emph{Forward-backward multiplicity correlations
in pp collisions in ALICE at 0.9, 2.76 and 7 TeV},
\emph{PoS} \textbf{Baldin ISHEPP XXI} (2012) 075

\end{thebibliography}
\end{document}